\documentstyle[12pt]{article}
\makeatletter
\def\abstract#1{\long\def\@abstract{#1}}%
\def\@abstract{}%
\let\@oldmaketitle=\@maketitle%
\def\@maketitle{%
\@oldmaketitle%
\begin{center}\large\bf Abstract\end{center}%
\begin{quotation}\@abstract\end{quotation}%
\vskip 1.5em}%
\makeatother
\makeatletter
\def\thesection{\arabic{section}}
\@addtoreset{equation}{section}
\def\theequation{\thesection.\arabic{equation}}
\def\appendix{\par
\setcounter{section}{0}
\setcounter{subsection}{0}
\def\thesection{\Alph{section}}}
\makeatother
\makeatletter
\def\eqnarray{%
\stepcounter{equation}%
\let\@currentlabel=\theequation
\global\@eqnswtrue
\global\@eqcnt\z@
\tabskip\@centering
\let\\=\@eqncr
$$\halign to \displaywidth\bgroup\@eqnsel\hskip\@centering
$\displaystyle\tabskip\z@{##}$&\global\@eqcnt\@ne
\hfil$\displaystyle{{}##{}}$\hfil
&\global\@eqcnt\tw@$\displaystyle\tabskip\z@{##}$\hfil
\tabskip\@centering&\llap{##}\tabskip\z@\cr}
\makeatother
\makeatletter
\@definecounter{whichequation}
\def\themanyeqn{\theequation-\alph{whichequation}}
\def\@manyeqnnum{{\rm (\themanyeqn)}}

\def\manyeqns{\stepcounter{equation}\let\@currentlabel=\themanyeqn

 \setcounter{whichequation}{1}
\global\@eqnswtrue
\global\@eqcnt\z@\tabskip\@centering\let\\=\@manyeqncr
$$\halign to \displaywidth\bgroup\@eqnsel\hskip\@centering
  $\displaystyle\tabskip\z@
 \let\@currentlabel=\theequation
  {##}$&\global\@eqcnt\@ne
  \hskip 2\arraycolsep \hfil${##}$\hfil
  &\global\@eqcnt\tw@ \hskip 2\arraycolsep $\displaystyle\tabskip\z@{##}$\hfil
   \tabskip\@centering&\llap{##}\tabskip\z@\cr}

\def\endmanyeqns{\@@manyeqncr\egroup
    \global\advance\c@equation\m@ne$$\global\@ignoretrue
    \stepcounter{equation}}

\def\@manyeqncr{{\ifnum0=`}\fi\@ifstar{\global\@eqpen\@M
    \@ymanyeqncr}{\global\@eqpen\interdisplaylinepenalty \@ymanyeqncr}}

\def\@ymanyeqncr{\@ifnextchar [{\@xmanyeqncr}{\@xmanyeqncr[\z@]}}

\def\@xmanyeqncr[#1]{\ifnum0=`{\fi}\@@manyeqncr
   \noalign{\penalty\@eqpen\vskip\jot\vskip #1\relax}}

\def\@@manyeqncr{\let\@tempa\relax
    \ifcase\@eqcnt \def\@tempa{& & &}\or \def\@tempa{& &}
      \else \def\@tempa{&}\fi
 \@tempa \if@eqnsw\@manyeqnnum\stepcounter{whichequation}\fi
     \global\@eqnswtrue\global\@eqcnt\z@\cr}
\makeatother
\newcommand{\fxi}{\mbox{\boldmath$\xi$}}

\newcommand{\ffm}{\mbox{\boldmath$m$}}
\newcommand{\ffl}{\mbox{\boldmath$l$}}

\newcommand{\ffp}{\mbox{\boldmath$\varphi$}}
\newcommand{\fft}{\mbox{\boldmath$\theta$}}

\newcommand{\delphi}{\Delta\varphi}

\newcommand{\deltheta}{\Delta\theta}
\newcommand{\delt}{\Delta t}

\newcommand{\measure}[2]{{d\mu\!\left({#1},{#2}\right)}}
\newcommand{\bra}[1]{{\langle{#1}\vert}}
\newcommand{\ket}[1]{{\vert{#1}\rangle}}
\newcommand{\braket}[2]{{\langle{#1}\vert{#2}\rangle}}
\newcommand{\kansu}[2]{{{#1}\!\left({#2}\right)}}
\newcommand{\kakko}[1]{{\left({#1}\right)}}

\begin{document}

\title{\sl \vspace{-25mm}
    Multi-Periodic Coherent States\\ and\\ the WKB-Exactness}

\author{
  Kazuyuki FUJII\thanks{e-mail address : fujii@yokohama-cu.ac.jp}\\
  Department of Mathematics, Yokohama City University,\\
  Yokohama 236, Japan\\\\
  Kunio FUNAHASHI\thanks{e-mail address : fnhs1scp@mbox.nc.kyushu-u.ac.jp}\\
  Department of Physics, Kyushu University,\\
  Fukuoka 812-81, Japan}
\date{May 1, 1996}

\abstract{
We construct the path integral formula in terms of
``multi-periodic'' coherent state as an extension of the
Nielsen-Rohrlich formula for spin.
We make an exact calculation of the formula and show that,
when a parameter corresponding to the magnitude of spin
becomes large, the leading order term of the expansion
coincides with the exact result.
We also give an explicit correspondence between the trace
formula in the multi-periodic coherent state and
the one in the ``generalized'' coherent state.
}
\maketitle\thispagestyle{empty}
\newpage

\section{Introduction}\label{sec:jo}

Spin, system being subject to the $SU(2)$ group, is a simple
and pedagogical model so that many useful results and
discussions have been made.
In the framework of the path integral, it is applied to,
for example, the explanation of the Fermi-Bose transmutation
\cite{APo,AS,TJ}.

In quantum mechanics there exist few systems which are
solved exactly so that various approximations such as
the perturbation or the variational method have been invented
to give useful results.
The most suitable one in the path integral formalism
is the WKB approximation, which can be stated as the stationary
phase method in quantum mechanics.
Although the WKB approximation is useful, the result differs from
the exact one in general.
However it is known that there exist some systems in which
they coincide with each other, for example, the harmonic oscillator:
this is a kind of trivial example since the approximation
gives the same form with the original action which is the
Gaussian to be integrated exactly.

In the stationary phase approximation in finite dimensions,
the conditions that an approximation gives the exact result
of the integral has been discovered by Duistermaat and
Heckman~\cite{DH,Atiyah},
so that the fact is called as the DH theorem.

Recently there have been some
discussions~\cite{Stone,Rajeev,Picken,Blau}
that the WKB approximation gives the exact result
(we call this fact as the {\bf WKB-exactness})
in the trace formula of spin in connection with the
DH theorem.
However there are some unsatisfactory points
in the preceding discussions,
for example, the manner of construction of the trace formula
and the use of the continuum path integral.
With paying attention to the above points, we have shown the
WKB-exactness in terms of the spin coherent state~\cite{FKSF1}
with the form
\begin{equation}\label{jo:trscs}
  Z =
  \lim_{N\to\infty}\left.
    \int\prod^N_{i=1}\measure{\xi_i}{\xi_i^*}
    \exp\left[ iJ\sum^N_{k=1}
      \left\{2i\kansu{\log}
        {{1+\xi_k^*\xi_k\over1+\xi_k^*\xi_{k-1}}}
        +\delt h{1-\xi_k^*\xi_{k-1}
          \over1+\xi_k^*\xi_{k-1}}\right\}\right]
  \right\vert_{\xi_N=\xi_0}\ ,
  \nonumber
\end{equation}
where $J$ is the magnitude of spin and $h$ the external magnetic
field.
We have also extended the WKB-exactness in terms of
the ``generalized'' coherent state~\cite{FKSF2}.

As for path integrals for spin, there exists another expression,
the Nielsen-Rohrlich formula~\cite{NR,Kashiwa}, which is constructed
in terms of the periodic coherent state.
Its trace formula is
\begin{eqnarray}\label{jo:nrf}
Z
&=&\sum^\infty_{n=-\infty}
\lim_{N\to\infty}\int^{2\kakko{n+1}\pi}_{2n\pi}
{d\varphi_N\over2\pi}
\int^\infty_{-\infty}\prod^{N-1}_{i=1}{d\varphi_i\over2\pi}
\int^{J+1/2-\varepsilon}_{-J-1/2+\varepsilon}
\prod^N_{j=1}{dp_j\over2\pi}
\\
&&\left.\times\exp\left[ i\sum^N_{k=1}
\kakko{p_k+J}\kakko{\varphi_k-\varphi_{k-1}}
-\delt\sum^N_{k=1}\kansu{h}{k}p_k\right]
\right\vert_{\varphi_N=\varphi_0+2n\pi}
\ ,\nonumber
\end{eqnarray}
where $\varepsilon$ is positive infinitesimal.
On the other hand, the trace formula of the spin coherent state
is (\ref{jo:trscs}).
The two representations look quite different despite
starting from the same Hamiltonian.
The integration domain of $p$'s and the existence of
infinite sum lead to the observation that the phase space is
considered  to be the ``punctured sphere''.
While the phase space of the spin coherent state is
$CP^1\simeq S^2$, the ``sphere''.
In the latter case, the WKB-exactness holds.
However in the former case, from (\ref{jo:nrf}), the equations of motion are
\begin{equation}\label{jo:eqm}
\cases{
  \displaystyle\varphi_k-\varphi_{k-1}=\kansu{h}{k}\delt\ ,\cr
  \displaystyle p_k-p_{k-1}=0\ ,\cr
}
\end{equation}
which does not meet the boundary condition $\varphi_N=\varphi_0+2n\pi$
in a general $\kansu{h}{k}$.
Thus there seems to be no classical solutions.
We have clarified this puzzle in ref. \cite{FKNS}.
The phase space of the Nielsen-Rohrlich formula is not ``punctured''
and the appearance of the infinite sum is superficial.
The trace formula is equivalent to that of the spin coherent state.
By rewriting (\ref{jo:nrf}) to
\begin{eqnarray}\label{jo:nrformulatheta}
Z &=& \sum^\infty_{n=-\infty}e^{i2n\pi J}\lim_{N\to\infty}
\int^{2\kakko{n+1}\pi}_{2n\pi}{d\varphi_N\over2\pi}
\int^\infty_{-\infty}\prod^{N-1}_{i=1}{d\varphi_i\over2\pi}
\int^\pi_0\prod^N_{j=1}\lambda\sin\theta_jd\theta_j
\nonumber\\
&&\times\exp\left\{ i\lambda\sum^N_{k=1}\cos\theta_k
\kakko{\varphi_k-\varphi_{k-1}-h\delt}\right\}
\Bigg\vert_{\varphi_N=\varphi_0+2n\pi}\ ,
\end{eqnarray}
the WKB-exactness also holds in this case.

Although the extension of the Nielsen-Rohrlich formula has been
attempted~\cite{Alekseev}, the explicit form has not been yet obtained.
We in this paper construct the path integral formula and clarify
the WKB-exactness.

The contents of this paper are as follows.
In Section \ref{sec:kousei}, we define the ``multi-periodic''
coherent state to construct the path integral formula.
In Section \ref{sec:genmitsu}, we make an exact calculation
of the trace formula.
We establish a relationship with the trace formula to that of the
generalized coherent state in Section \ref{sec:multi:relationship}.
Then we perform the WKB approximation to confirm the WKB-exactness
in Section \ref{sec:multi:wkb}.
The last section is devoted to the discussion.

\section{Construction of the Trace Formula}\label{sec:kousei}

First we make up the $u(N+1)$ algebra and the representation
space by means of $N+1$ harmonic oscillators.
The method is called the Schwinger boson one~\cite{Schwinger}.
The commutation relations of oscillators are
\begin{equation}\label{harm:daisu}
  \big[ a_\alpha,a^\dagger_\beta\big] = \delta_{\alpha\beta}\ ,\
  \big[ a_\alpha,a_\beta\big] =
  \big[ a^\dagger_\alpha,a^\dagger_\beta\big] = 0
  \ ,\ \kakko{\alpha,\beta=1,\cdots N+1}\ ,
\end{equation}
and the Fock space is
\begin{eqnarray}\label{cpn:fock}
  &&\left\{\ket{m^1,\cdots,m^{N+1}}\right\}\ ,\
  \kakko{m^\alpha=0,1,2,\cdots\ {\rm with}\ \alpha=1,\cdots,N+1}\ ,
  \nonumber\\
  &&\ket{m^1,\cdots,m^{N+1}}\equiv
  {1\over\sqrt{m^1!\cdots m^{N+1}!}}
  \kakko{a^\dagger_1}^{m^1}\cdots
  \kakko{a^\dagger_{N+1}}^{m^{N+1}}\ket{0}\ ,
  \nonumber\\
  &&a_\alpha\ket{0} = 0\ .
\end{eqnarray}

By putting
\begin{equation}
  E_{\alpha\beta}=a^\dagger_\alpha a_\beta\ ,\
  \kakko{\alpha,\beta=1,\cdots,N+1}\ ,
\end{equation}
the $u(N+1)$ algebra
\begin{eqnarray}
  \big[ E_{\alpha\beta},E_{\gamma\delta}\big] =
  \delta_{\beta\gamma}
  E_{\alpha\delta}-\delta_{\delta\alpha}E_{\gamma\beta}\ ,
\end{eqnarray}
is realized.

The Fock space (\ref{cpn:fock}) is too large to be the representation
space of $U(N+1)$.
We restrict the representation space with the identity operator
\begin{equation}\label{cpn:ichiq}
  {\bf 1}_Q\equiv\sum_{\| m\|=Q}
  \ket{m^1,\cdots,m^{N+1}}\bra{m^1,\cdots,m^{N+1}}\ ,
\end{equation}
where we have used the abbreviation
\begin{equation}
  \| m\| \equiv \sum^{N+1}_{\alpha=1}m^\alpha\ .
\end{equation}

Now we introduce the highest weight vector defined by
\begin{eqnarray}
  &&E_{N+1,N+1}\vert Q;N+1\rangle\rangle =
  Q\vert Q;N+1\rangle\rangle\ ,\
  \nonumber\\
  &&E_{N+1,\alpha}\vert Q;N+1\rangle\rangle = 0\ ,\
  \kakko{\alpha=1,\cdots,N}\ ,
\end{eqnarray}
where $E_{\alpha,N+1}(E_{N+1,\alpha})$ is the lowering (raising)
operator of $u(N+1)$.
We can identify
\begin{equation}\label{cpn:highest}
  \vert Q;N+1\rangle\rangle \equiv
  \ket{\stackrel{1}{0},\cdots,\stackrel{N}{0},\stackrel{N+1}{Q}}\ .
\end{equation}

The ``multi-periodic'' coherent state is defined by
\begin{eqnarray}
  \ket{\ffp} &\equiv&
  \ket{\varphi^1,\varphi^2,\cdots,\varphi^N}
  \nonumber\\
  &\equiv&
  {1\over\kakko{2\pi}^{N/2}}\sum_{\| m\|=Q}\prod^N_{\alpha=1}
  \kakko{e^{-im^\alpha\varphi^\alpha}
    \kakko{E_{\alpha,N+1}}^{m^\alpha}}
  \vert Q;N+1\rangle\rangle
  \nonumber\\
  &=&
  {1\over\kakko{2\pi}^{N/2}}\sum_{\| m\|=Q}
  e^{-i\sum^N_{\alpha=1}m^\alpha\varphi^\alpha}
  \ket{m^1,m^2,\cdots,m^{N+1}}\ ,
\end{eqnarray}
which satisfies
periodicity
\begin{equation}
  \ket{\varphi^1,\cdots,\varphi^\alpha+2n\pi,\cdots,\varphi^N} =
  \ket{\varphi^1,\cdots,\varphi^\alpha,\cdots,\varphi^N}\ ,\
  \kakko{n:{\rm integer}}\ .
\end{equation}
The inner product is
\begin{equation}\label{multi:naiseki}
  \braket{\ffp}{\ffp^\prime} =
  {1\over\kakko{2\pi}^N}\sum_{\| m\|=Q}
  e^{i\sum^N_{\alpha=1}m^\alpha
    \kakko{\varphi^\alpha-{\varphi^\alpha}^\prime}}\ ,
\end{equation}
and the resolution of unity
\begin{equation}\label{multi:resofuni}
  \int^{2\pi}_0d\ffp\ket{\ffp}\bra{\ffp} = {\bf 1}_Q\ ,
\end{equation}
where
\begin{equation}
  \int^{2\pi}_0d\ffp \equiv
  \int^{2\pi}_0\prod^N_{\alpha=1}
  d\varphi^\alpha\ .
\end{equation}

To construct the path integral formula, we utilize the next relation:
\begin{eqnarray}\label{nr:kankeishiki}
  &&{\rm for}\ m_0,m_1\in{\bf Z}
  \nonumber\\
  &&\sum^{m_1}_{m=m_0}
  e^{im\varphi}\kansu{f}{m}
  =\sum^\infty_{n=-\infty}
  \int^{m_1+\varepsilon}_{m_0-\varepsilon}
  dpe^{ip\kakko{\varphi+2n\pi}}
  \kansu{f}{p}\ ,\
  \kakko{0<\varepsilon<1}\ .
\end{eqnarray}
We rewrite (\ref{multi:naiseki}) with the aid of (\ref{nr:kankeishiki}) to
\begin{eqnarray}\label{multi:naisekinakaba}
\braket{\ffp}{\ffp^\prime}
&=&
{1\over\kakko{2\pi}^N}\sum^Q_{m^1=0}\sum^{Q-m^1}_{m^2=0}
\cdots\sum^{Q-\sum^{N-1}_{\alpha=1}m^\alpha}_{m^N=0}
e^{i\sum^N_{\alpha=1}m^\alpha
\kakko{\varphi^\alpha-{\varphi^\alpha}^\prime}}
\nonumber\\
&=&
{1\over\kakko{2\pi}^N}\sum^Q_{m^1=0}\sum^{Q-m^1}_{m^2=0}
\cdots\sum^{Q-\sum^{N-2}_{\alpha=1}m^\alpha}_{m^{N-1}=0}
e^{i\sum^N_{\alpha=1}m^\alpha
\kakko{\varphi^\alpha-{\varphi^\alpha}^\prime}}
\nonumber\\
&&\times
\sum^\infty_{n^N=-\infty}
\int^{Q-\sum^{N-1}_{\alpha=1}m^\alpha
+\varepsilon_N}_{-\varepsilon_N}
dp^Ne^{ip^N\kakko{\varphi^N-{\varphi^N}^\prime+2n^N\pi}}
\nonumber\\
&=&\cdots
\nonumber\\
&=&
{1\over\kakko{2\pi}^N}
\sum^\infty_{n^1=-\infty}\cdots\sum^\infty_{n^N=-\infty}
\int^{Q+\varepsilon_1}_{-\varepsilon_1}dp^1
\int^{Q-p^1+\varepsilon_2}_{-\varepsilon_2}dp^2
\nonumber\\
&&\times\cdots
\int^{Q-\sum^{N-2}_{\alpha=1}p^\alpha
+\varepsilon_{N-1}}_{-\varepsilon_{N-1}}dp^{N-1}
\int^{Q-\sum^{N-1}_{\alpha=1}p^\alpha
+\varepsilon_N}_{-\varepsilon_N}dp^N
\nonumber\\
&&\times e^{i\sum^N_{\alpha=1}p^\alpha
\kakko{\varphi^\alpha-{\varphi^\alpha}^\prime+2n^\alpha\pi}}\ ,
\end{eqnarray}
where we restrict $0<\varepsilon_\alpha\le1/\kakko{2^{\alpha-1}\kakko{N+1}}$
for later convenience.
We make a change of variables such that
\begin{equation}
\tilde p^\alpha = p^\alpha+{1\over2^{\alpha-1}\kakko{N+1}}\ ,
\end{equation}
with putting
\begin{eqnarray}\label{multi:hani}
\lambda &\equiv& {Q\over2}+{1\over N+1}\ ,
\nonumber\\
\tilde\varepsilon_\alpha &\equiv&
-\varepsilon_\alpha+{1\over2^{\alpha-1}\kakko{N+1}}\ ,\
\kakko{0\le\tilde\varepsilon_\alpha<{1\over2^{\alpha-1}\kakko{N+1}}}\ ,
\end{eqnarray}
which leads to
\begin{equation}
\int^{Q-\sum^{\alpha-1}_{\beta=1}p^\beta+\varepsilon_\alpha}
_{-\varepsilon_\alpha}dp^\alpha\to
\int^{2\lambda-\sum^{\alpha-1}_{\beta=1}\tilde p^\beta
-\tilde\varepsilon_\alpha}
_{\tilde\varepsilon_\alpha}d\tilde p^\alpha\ .
\end{equation}
 (\ref{multi:naisekinakaba}) then becomes
\begin{eqnarray}
\braket{\ffp}{\ffp^\prime}
&=&
{1\over\kakko{2\pi}^N}
\sum^\infty_{n^1=-\infty}\cdots\sum^\infty_{n^N=-\infty}
\int^{2\lambda-\varepsilon_1}_{\varepsilon_1}dp^1
\int^{2\lambda-p^1-\varepsilon_2}_{\varepsilon_2}dp^2\cdots
\nonumber\\
&&\times
\int^{2\lambda-\sum^{N-2}_{\alpha=1}p^\alpha-\varepsilon_{N-1}}
_{\varepsilon_{N-1}}dp^{N-1}
\int^{2\lambda-\sum^{N-1}_{\alpha=1}p^\alpha-\varepsilon_N}
_{\varepsilon_N}dp^N
\nonumber\\
&&\times
e^{i\sum^N_{\alpha=1}\kakko{p^\alpha-{1\over2^{\alpha-1}\kakko{N+1}}}
\kakko{\varphi^\alpha-{\varphi^\alpha}^\prime+2n^\alpha\pi}}\ ,
\end{eqnarray}
where the tildes have been omitted.

We adopt a Hamiltonian as
\begin{equation}\label{cpn:hamiltonian}
  \hat H \equiv
  \sum^{N+1}_{\alpha=1}c_\alpha a_\alpha^\dagger a_\alpha\ ,
\end{equation}
where we have assumed that all $c$'s are different from each other and
independent on time.
The matrix element of the Hamiltonian is
\begin{eqnarray}
\bra{\ffp}\hat H\ket{\ffp^\prime}
&=&
\left\{\sum^N_{\alpha=1}\kakko{c_\alpha-c_{N+1}}{1\over i}
{\partial\over\partial\varphi^\alpha}
+c_{N+1}Q\right\}\braket{\ffp}{\ffp^\prime}
\nonumber\\
&=&
{1\over\kakko{2\pi}^N}
\sum^\infty_{n^1=-\infty}\cdots\sum^\infty_{n^N=-\infty}
\int^{2\lambda-\varepsilon_1}_{\varepsilon_1}dp^1\cdots
\int^{2\lambda-\sum^{N-1}_{\alpha=1}p^\alpha
-\varepsilon_N}_{\varepsilon_N}dp^N
\nonumber\\
&&\times
e^{i\sum^N_{\alpha=1}\kakko{p^\alpha-{1\over2^{\alpha-1}\kakko{N+1}}}
\kakko{\varphi^\alpha-{\varphi^\alpha}^\prime+2n^\alpha\pi}}
\nonumber\\
&&\times
\left[\sum^N_{\alpha=1}\mu_\alpha
\kakko{p^\alpha-{1\over2^{\alpha-1}\kakko{N+1}}}+c_{N+1}Q\right]\ ,
\end{eqnarray}
where $\mu_\alpha\equiv c_\alpha-c_{N+1}$.

{}From now on we put $\tilde\varepsilon_\alpha=0$ for simplicity.
(See (\ref{multi:hani}) for the range of $\tilde\varepsilon_\alpha$.)

The Feynman kernel is defined by
\begin{eqnarray}
  \kansu{K}{\ffp_F,\ffp_I;T} \equiv
  \bra{\ffp_F}e^{-i\hat HT}\ket{\ffp_I} =
  \lim_{M\to\infty}\bra{\ffp_F}
  \kakko{1-i\delt\hat H}^M\ket{\ffp_I}\ ,
  \nonumber\\
  \kakko{\delt\equiv T/M}\ ,
\end{eqnarray}
or explicitly
\begin{eqnarray}
  \kansu{K}{\ffp_F,\ffp_I;T}
  &=&
  \lim_{M\to\infty}\int^{2\pi}_0
  \prod^{M-1}_{i=1}d\ffp_i
  \prod^M_{j=1}\bra{\ffp_j}
  \kakko{1-i\delt\hat H}\ket{\ffp_{j-1}}
  \Bigg\vert^{\ffp_M=\ffp_F}_{\ffp_0=\ffp_I}
  \nonumber\\
  &=&
  \lim_{M\to\infty}\int^{2\pi}_0
  \prod^{M-1}_{i=1}d\ffp_i
  \prod^M_{j=1}\Bigg[{1\over\kakko{2\pi}^N}
  \sum^\infty_{n_j^1=-\infty}\cdots
  \sum^\infty_{n_j^N=-\infty}
  \nonumber\\
  &&\times
  \int^{2\lambda}_0dp_j^1\cdots
  \int^{2\lambda-\sum^{N-1}_{\alpha=1}p_j^\alpha}_0dp_j^N
  e^{i\sum^N_{\alpha=1}
    \kakko{p^\alpha-{1\over2^{\alpha-1}\kakko{N+1}}}
    \kakko{\delphi_j^\alpha+2n_j^\alpha\pi}}
  \nonumber\\
  &&\times
  \Bigg[1-i\delt\Bigg\{\sum^N_{\alpha=1}\mu_\alpha
  \kakko{p_j^\alpha-{1\over2^{\alpha-1}\kakko{N+1}}}
  +c_{N+1}Q\Bigg\}\Bigg]\Bigg]
  \Bigg\vert^{\ffp_M=\ffp_F}_{\ffp_0=\ffp_I}
  \nonumber\\
  &=&
  \lim_{M\to\infty}\int^{2\pi}_0
  \prod^{M-1}_{i=1}d\ffp_i
  \prod^M_{j=1}\left\{{1\over\kakko{2\pi}^N}
    \sum^\infty_{n_j^1=-\infty}\cdots
    \sum^\infty_{n_j^N=-\infty}\right.
  \nonumber\\
  &&\left.\times
    \int^{2\lambda}_0dp_j^1\cdots
    \int^{2\lambda-\sum^{N-1}_{\alpha=1}
      p_j^\alpha}_0dp_j^N\right\}
  \nonumber\\
  &&\times
  \exp\left[ i\sum^M_{k=1}\left\{\sum^N_{\alpha=1}
      \kakko{p_k^\alpha-{1\over2^{\alpha-1}\kakko{N+1}}}
      \kakko{\delphi_k^\alpha+2n_k^\alpha\pi}\right.\right.
  \nonumber\\
  &&\left.\left.\left.
        -\delt\left\{\sum^N_{\alpha=1}\mu_\alpha
          \kakko{p_k^\alpha-{1\over2^{\alpha-1}\kakko{N+1}}}
          +c_{N+1}Q\right\}\right\}\right]
  \right\vert^{\ffp_M=\ffp_F}_{\ffp_0=\ffp_I}\ ,
\end{eqnarray}
where the resolution of unity (\ref{multi:resofuni})
has been inserted in the first equality and
$\delphi_k^\alpha\equiv\varphi_k^\alpha-\varphi_{k-1}^\alpha$
has been put and $\kansu{\rm O}{(\delt)^2}$ terms,
which finally vanish in $M\to\infty$ limit, have been omitted
in the last equality.
We introduce new variables such that
\begin{equation}
  \cases{
    \displaystyle
    {n_k^\alpha}^\prime =
    \sum^k_{l=1}n_l^\alpha\ ,\cr
    \displaystyle
    {\varphi_k^\alpha}^\prime =
    \varphi_k^\alpha+2{n_k^\alpha}^\prime\pi\ ,\cr}
\end{equation}
to give
\begin{equation}
  \cases{
    \displaystyle
    \delphi_k^\alpha+2n_k^\alpha\pi =
    {\delphi_k^\alpha}^\prime\ ,\cr
    \displaystyle
    \sum^\infty_{{n_i^\alpha}^\prime=-\infty}
    \int^{2\kakko{{n_i^\alpha}^\prime+1}\pi}_{2{n_i^\alpha}^\prime\pi}
    d{\varphi_i^\alpha}^\prime =
    \int^\infty_{-\infty}d{\varphi_i^\alpha}^\prime\ .\cr}
\end{equation}
The kernel is then rewritten
\begin{eqnarray}
  \kansu{K}{\ffp_F,\ffp_I;T}
  &=&
  \lim_{M\to\infty}\int^\infty_{-\infty}
  \prod^{M-1}_{i=1}d\ffp_i
  \sum^\infty_{n_M^1=-\infty}\cdots
  \sum^\infty_{n_M^N=-\infty}
  \nonumber\\
  &&\times
  \prod^M_{j=1}\left\{
    \int^{2\lambda}_0{dp_j^1\over2\pi}\cdots
    \int^{2\lambda-\sum^{N-1}_{\alpha=1}p_j^\alpha}_0
    {dp_j^N\over2\pi}\right\}
  \nonumber\\
  &&\times
  \exp\left[ i\sum^M_{k=1}\left\{\sum^N_{\alpha=1}
      \kakko{p_k^\alpha-{1\over2^{\alpha-1}\kakko{N+1}}}
      \delphi_k^\alpha\right.\right.
  \nonumber\\
  &&\left.\left.\left.
        -\delt\left\{\sum^N_{\alpha=1}\mu_\alpha
          \kakko{p_k^\alpha-{1\over2^{\alpha-1}\kakko{N+1}}}
          +c_{N+1}Q\right\}\right\}\right]
  \right\vert^{\varphi^\alpha_M=\varphi^\alpha_F+2n_M^\alpha\pi}
  _{\varphi^\alpha_0=\varphi^\alpha_I}
  \nonumber\\
  &=&
  e^{-iQc_{N+1}T+i\sum^N_{\alpha=1}
    {1\over2^{\alpha-1}\kakko{N+1}}\mu_\alpha T}
  \nonumber\\
  &&\times
  \sum^\infty_{n^1=-\infty}\cdots
  \sum^\infty_{n^N=-\infty}
  e^{-i\sum^N_{\alpha=1}{1\over2^{\alpha-1}\kakko{N+1}}
    \kakko{\varphi_F^\alpha-\varphi_I^\alpha+2n^\alpha\pi}}
  \nonumber\\
  &&\times
  \lim_{M\to\infty}\int^\infty_{-\infty}
  \prod^{M-1}_{i=1}d\ffp_i
  \prod^M_{j=1}\left\{\int^{2\lambda}_0{dp_j^1\over2\pi}\cdots
    \int^{2\lambda-\sum^{N-1}_{\alpha=1}p_j^\alpha}_0
    {dp_j^N\over2\pi}\right\}
  \nonumber\\
  &&\times
    \exp\Bigg[ i\sum^M_{k=1}\sum^N_{\alpha=1}p_k^\alpha
    \Phi^\alpha_k\Bigg]
  \Bigg\vert^{\varphi^\alpha_M=\varphi^\alpha_F+2n^\alpha\pi}
  _{\varphi^\alpha_0=\varphi^\alpha_I}\ ,
\end{eqnarray}
where primes have been omitted and $n_M^\alpha$ has been written as
$n^\alpha$ and $\Phi^\alpha_k\equiv\delphi^\alpha_k-\delt\mu_\alpha$
in the last equality.

Further we make a change of variables such that
\begin{eqnarray}
  {p^1}^\prime &=& p^1-\lambda\ ,
  \nonumber\\
  {p^2}^\prime &=& p^2-{\lambda-{p^1}^\prime\over2}\ ,
  \nonumber\\
  &\vdots&
  \nonumber\\
  {p^\alpha}^\prime &=&
  p^\alpha-{\lambda-\sum^{\alpha-1}_{\beta=1}
    2^{\beta-1}{p^\beta}^\prime\over2^{\alpha-1}}\ ,\
  \nonumber\\
  &\vdots&
  \nonumber\\
  {p^N}^\prime &=&
  p^N-{\lambda-\sum^{N-1}_{\beta=1}
    2^{\beta-1}{p^\beta}^\prime\over2^{N-1}}\ ,\
\end{eqnarray}
which leads to
\begin{equation}
  \int^{2\lambda-\sum^{\alpha-1}_{\beta=1}p^\beta}_0
  dp^\alpha\longrightarrow
  \int^{(\lambda-\sum^{\alpha-1}_{\beta=1}
    2^{\beta-1}{p^\beta}^\prime)/2}
  _{-(\lambda-\sum^{\alpha-1}_{\beta=1}
    2^{\beta-1}{p^\beta}^\prime)/2}
  d{p^\alpha}^\prime\ ,
\end{equation}
where we use the notations,
\begin{equation}
  \cases{
    \displaystyle \sum^b_{\beta=a}\kansu{f}{\beta}=0\cr
    \displaystyle \prod^b_{\beta=a}\kansu{f}{\beta}=1\cr
    }\ ,\
  \kakko{{\rm for}\ b<a}\ ,
\end{equation}
for simplicity.
We then obtain the kernel
\begin{eqnarray}
  \kansu{K}{\ffp_F,\ffp_I;T}
  &=&
  e^{-iQc_{N+1}T+i\sum^N_{\alpha=1}
    {1\over2^{\alpha-1}\kakko{N+1}}\mu_\alpha T}
  \nonumber\\
  &&\times
  \sum^\infty_{n^1=-\infty}\cdots
  \sum^\infty_{n^N=-\infty}
  e^{-i\sum^N_{\alpha=1}
    {1\over2^{\alpha-1}\kakko{N+1}}
    \kakko{\varphi^\alpha_F
      -\varphi^\alpha_I+2n^\alpha\pi}}
  \nonumber\\
  &&\times
  \lim_{M\to\infty}
  \int^\infty_{-\infty}\prod^{M-1}_{i=1}
  d\ffp_i\prod^M_{j=1}
  \left\{\int^{\lambda}_{-\lambda}{dp_j^1\over2\pi}
    \int^{(\lambda-p^1_j)/2}_{-(\lambda-p^1_j)/2}
    {dp_j^2\over2\pi}\right.
  \nonumber\\
  &&\left.\times\cdots
    \int^{(\lambda-\sum^{N-1}_{\beta=1}2^{\beta-1}
      p_j^\beta)/2^{N-1}}
    _{-(\lambda-\sum^{N-1}_{\beta=1}2^{\beta-1}
      p_j^\beta)/2^{N-1}}
    {dp^N_j\over2\pi}\right\}
  \\
  &&\left.\times
    \exp\left[ i\sum^M_{k=1}\sum^N_{\alpha=1}
      \kakko{p_k^\alpha+{\lambda
          -\sum^{\alpha-1}_{\beta=1}2^{\beta-1}
          p_k^\beta\over2^{\alpha-1}}}
      \Phi^\alpha_k\right]
  \right\vert^{\varphi^\alpha_M =
    \varphi^\alpha_F+2n^\alpha\pi}
  _{\varphi^\alpha_0=\varphi^\alpha_I}\ ,
  \nonumber
\end{eqnarray}
where we have again omitted primes.

To compare with the $\theta$-expression of the Nielsen-Rohrlich
formula (\ref{jo:nrformulatheta}),
we make a change of variables such that
\begin{equation}
 p^\alpha_k =
 \lambda\kakko{\prod^{\alpha-1}_{\beta=1}
   \sin^2{\theta_k^\beta\over2}}
 \cos\theta_k^\alpha\ ,\
 \kakko{\alpha=1,\cdots,N}\ ,
\end{equation}
with the Jacobian
\begin{equation}
  {\kansu{\partial}{p}\over\kansu{\partial}{\theta}} =
  \kakko{-1}^N\lambda^N\prod^N_{\alpha=1}
  \Bigg\{\kakko{\sin^2{\theta^\alpha_k\over2}}
    ^{N-\alpha}\sin\theta^\alpha_k\Bigg\}\ ,
\end{equation}
to obtain
\begin{eqnarray}
  &&\hspace{-10mm}\kansu{K}{\ffp_F,\ffp_I;T}
  =
  e^{-iQc_{N+1}T+i\sum^N_{\alpha=1}
    {1\over2^{\alpha-1}\kakko{N+1}}\mu_\alpha T}
  \nonumber\\
  &&\times
  \sum^\infty_{n^1=-\infty}\cdots\sum^\infty_{n^N=-\infty}
  e^{-i\sum^N_{\alpha=1}{1\over2^{\alpha-1}\kakko{N+1}}
    \kakko{\varphi_F^\alpha-\varphi_I^\alpha+2n^\alpha\pi}}
  \nonumber\\
  &&\times
  \lim_{M\to\infty}\int^\infty_{-\infty}\prod^{M-1}_{i=1}d\ffp_i
  \prod^M_{j=1}\left\{\lambda^N\int^\pi_0
    \kakko{\sin^2{\theta_j^1\over2}}^{N-1}
    \sin\theta_j^1{d\theta_j^1\over2\pi}\right.
  \nonumber\\
  &&\left.\times\cdots
    \int^\pi_0\sin^2{\theta_j^{N-1}\over2}\sin\theta_j^{N-1}
    {d\theta_j^{N-1}\over2\pi}
    \int^\pi_0\sin\theta_j^N{d\theta_j^N\over2\pi}\right\}
  \nonumber\\
  &&\times
  \exp\bigg[ i\sum^M_{k=1}\sum^N_{\alpha=1}2\lambda
    \kakko{\prod^{\alpha-1}_{\beta=1}
      \sin^2{\theta_k^\beta\over2}}
    \cos^2{\theta_k^\alpha\over2}
      \Phi^\alpha_k\bigg]
  \bigg\vert^{\varphi^\alpha_M=\varphi^\alpha_F+2n^\alpha\pi}
  _{\varphi^\alpha_0=\varphi^\alpha_I}\ .
\end{eqnarray}

The trace formula is defined by
\begin{equation}
  Z \equiv
  \int^{2\pi}_0d\ffp\bra{\ffp}
  e^{-i\hat HT}\ket{\ffp} =
  \int^{2\pi}_0d\ffp\kansu{K}{\ffp,\ffp;T}\ ,
\end{equation}
whose explicit form is
\begin{eqnarray}\label{multi:traceform}
  Z
  &=&
  e^{-iQc_{N+1}T+i{1\over N+1}
    \sum^N_{\alpha=1}2^{-\alpha+1}\mu_\alpha T}
  \nonumber\\
  &&\times
  \sum^\infty_{n^1=-\infty}\cdots
  \sum^\infty_{n^N=-\infty}
  e^{-i{1\over N+1}\sum^N_{\alpha=1}
    2^{-\alpha+2}n^\alpha\pi}
  \nonumber\\
  &&\times
  \lim_{M\to\infty}\prod^N_{\alpha=1}\left\{
    \int^{2\kakko{n^\alpha+1}\pi}_{2n^\alpha\pi}
    d\varphi_M^\alpha
  \right\}\int^\infty_{-\infty}\prod^{M-1}_{i=1}d\ffp_i
  \nonumber\\
  &&\times
  \prod^M_{j=1}\left\{\lambda^N\int^\pi_0
    \kakko{\sin^2{\theta_j^1\over2}}^{N-1}
    \sin\theta_j^1{d\theta_j^1\over2\pi}\right.
  \nonumber\\
  &&\left.\times\cdots
    \int^\pi_0
    \sin^2{\theta_j^{N-1}\over2}
    \sin\theta_j^{N-1}{d\theta_j^{N-1}\over2\pi}
    \int^\pi_0\sin\theta_j^N
    {d\theta_j^N\over2\pi}\right\}
  \nonumber\\
  &&\times\left.
    \exp\left[2i\lambda\sum^M_{k=1}\sum^N_{\alpha=1}
      \kakko{\prod^{\alpha-1}_{\beta=1}
        \sin^2{\theta_k^\beta\over2}}
      \cos^2{\theta_k^\alpha\over2}
      \Phi^\alpha_k\right]
  \right\vert_{\varphi^\alpha_0=\varphi^\alpha_M-2n^\alpha\pi}\ .
\end{eqnarray}
This is the counterpart of the $\theta$-expression of the
Nielsen-Rohrlich formula.
Actually, in $N=1$ case, by putting $Q=2J$ and $c_1=-c_2=h/2$,
(\ref{multi:traceform}) becomes
\begin{eqnarray}
  Z &=&
  \sum^\infty_{n=-\infty}
  e^{i2n\pi J}\lim_{M\to\infty}
  \int^{2\kakko{n+1}\pi}_{2n\pi}
  {d\varphi_M\over2\pi}
  \int^\infty_{-\infty}
  \prod^{M-1}_{i=1}
  {d\varphi_i\over2\pi}
  \prod^M_{j=1}\bigg\{\lambda
  \int^\pi_0\sin\theta_jd\theta_j\bigg\}
  \nonumber\\
  &&\times
  \exp\bigg[
  i\lambda\sum^M_{k=1}
  \cos\theta_k
  \kakko{\delphi_k-h\delt}\bigg]\ ,
\end{eqnarray}
which is just the Nielsen-Rohrlich formula (\ref{jo:nrformulatheta}).

\section{Exact Calculation}\label{sec:genmitsu}

To claim the WKB-exactness, we need the exact calculation
to be compared with the result of the WKB approximation.
We have already obtained the exact result in another expression
(the generalized coherent state)~\cite{FKSF2}.
However we make an exact calculation in this expression
to be self-contained.

By changing variables such that
\begin{equation}
  p^\alpha_k =
  \kakko{\prod^{\alpha-1}_{\beta=1}
    \sin^2\frac{\theta^\beta_k}{2}}
  \cos^2\frac{\theta^\alpha_k}{2}\ ,
\end{equation}
with the Jacobian
\begin{equation}
  \frac{\kansu{\partial}{p}}{\kansu{\partial}{\theta}} =
  \kakko{-1}^N2^{-N}
  \prod^N_{\alpha=1}\Bigg\{
  \kakko{\sin^2{\theta^\alpha_k\over2}}^{N-\alpha}
  \sin\theta^\alpha_k\Bigg\}\ ,
\end{equation}
the trace formula (\ref{multi:traceform}) becomes
\begin{eqnarray}\label{multi:ptrace}
  Z &=&
  e^{-iQc_{N+1}T+i\frac{1}{N+1}
    \sum^N_{\alpha=1}2^{-\alpha+1}\mu_\alpha T}
  \sum^\infty_{n^1=-\infty}\cdots\
  \sum^\infty_{n^N=-\infty}
  e^{-i\frac{1}{N+1}
    \sum^N_{\alpha=1}2^{-\alpha+1}\mu_\alpha T}
  \nonumber\\
  &&\times
  \lim_{M\to\infty}\prod^M_{i=1}
  \bigg\{\kakko{2\lambda}^N
  \int^1_0dp^1_i\cdots
  \int^{1-\sum^{N-1}_{\alpha=1}p^\alpha_i}_0
  dp^N_i\bigg\}
  \nonumber\\
  &&\times
  \prod^N_{\alpha=1}
  \bigg\{\int^{2\kakko{n^\alpha+1}\pi}_{2n^\alpha\pi}
  \frac{d\varphi^\alpha_M}{2\pi}
  \int^\infty_{-\infty}\prod^{M-1}_{j=1}
  \frac{d\varphi^\alpha_j}{2\pi}\bigg\}
  \prod^N_{\alpha=1}
  e^{2i\lambda\sum^M_{k=1}p^\alpha_k
    \Phi^\alpha_k}\ .
\end{eqnarray}
Rewriting the exponent to
\begin{equation}
  \sum^M_{k=1}p^\alpha_k\Phi^\alpha_k =
  \sum^{M-1}_{k=1}\kakko{p^\alpha_k-p^\alpha_{k-1}}\varphi^\alpha_k
  +\kakko{p^\alpha_M-p^\alpha_1}\varphi^\alpha_M
  +p^\alpha_12n^\alpha\pi-\delt\mu_\alpha\sum^M_{k=1}p^\alpha_k\ ,
\end{equation}
and performing the $\varphi_i$-integrals $(i=1,\cdots,N-1)$
as the $\delta$-functions, we obtain
\begin{eqnarray}\label{multi:phinochi}
  Z &=&
  e^{-iQc_{N+1}T+i{1\over N+1}
    \sum^N_{\alpha=1}2^{-\alpha+1}\mu_\alpha T}
  \sum^\infty_{n^1=-\infty}
  \cdots\sum^\infty_{n^N=-\infty}
  e^{-i{1\over N+1}
    \sum^N_{\alpha=1}2^{-\alpha+1}2n^\alpha\pi}
  \nonumber\\
  &&\times
  \lim_{M\to\infty}\prod^M_{i=1}
  \bigg\{\kakko{2\lambda}^N
  \int^1_0dp^1_i\cdots
  \int^{1-\sum^{N-1}_{\alpha=1}p^\alpha_i}_0dp^N_i\bigg\}
  \nonumber\\
  &&\times\prod^N_{\alpha=1}\bigg[
  \int^{2\kakko{n^\alpha+1}\pi}_{2n^\alpha\pi}
  {d\varphi^\alpha_M\over2\pi}
  e^{2i\lambda\left\{
      \kakko{p^\alpha_M-p^\alpha_1}\varphi^\alpha_M
      +p^\alpha_12n^\alpha\pi
      -\delt\mu_\alpha\sum^M_{k=1}p^\alpha_k\right\}}
  \nonumber\\
  &&\times
  \prod^{M-1}_{j=1}\bigg\{{1\over2\lambda}
  \kansu{\delta}{p^\alpha_j-p^\alpha_{j+1}}\bigg\}\bigg]
  \nonumber\\
  &=&
  e^{-iQc_{N+1}T+i{1\over N+1}
    \sum^N_{\alpha=1}2^{-\alpha+1}\mu_\alpha T}
  \sum^\infty_{n^1=-\infty}
  \cdots\sum^\infty_{n^N=-\infty}
  e^{-i{1\over N+1}
    \sum^N_{\alpha=1}2^{-\alpha+1}2n^\alpha\pi}
  \nonumber\\
  &&\times
  \kakko{2\lambda}^N\int^1_0dp^1_M\cdots
  \int^{1-\sum^{N-1}_{\alpha=1}p^\alpha_M}_0dp^N_M
  e^{2i\lambda\sum^N_{\alpha=1}
    \kakko{2n^\alpha\pi-\mu_\alpha T}p^\alpha_M}\ .
\end{eqnarray}
By applying the integral relation (\ref{sekibunkoushiki})
with $I=0$, $L=1$, $u=2\lambda$  and
$\Phi^\alpha=2n^\alpha\pi-\mu_\alpha T$,
(\ref{multi:phinochi}) becomes
\begin{eqnarray}\label{multi:waketa}
  Z &=&
  e^{-iQc_{N+1}T+i{1\over N+1}
    \sum^N_{\alpha=1}2^{-\alpha+1}\mu_\alpha T}
  \sum^\infty_{n^1=-\infty}\cdots\sum^\infty_{n^N=-\infty}
  e^{-i{1\over N+1}
    \sum^N_{\alpha=1}2^{-\alpha+1}2n^\alpha\pi}
  \nonumber\\
  &&\times\Bigg[\sum^N_{\alpha=1}
  {e^{i2\lambda\kakko{2n^\alpha\pi-\mu_\alpha T}}\over
    i\kakko{2n^\alpha\pi-\mu_\alpha T}
      \prod^N_{\beta=1\atop\beta\ne\alpha}
      i\left\{2\kakko{n^\alpha-n^\beta}\pi
        -\kakko{\mu_\alpha-\mu_\beta}T\right\}}
  \nonumber\\
  &&+
  \frac{\kakko{-1}^N}{\prod^N_{\alpha=1}i
    \kakko{2n^\alpha\pi-\mu_\alpha T}}\Bigg]\ .
\end{eqnarray}
First we calculate each term of $\alpha$-sum.
We set
\begin{eqnarray}\label{multi:kobetsu}
  Z_\alpha &\equiv&
  e^{-iQc_{N+1}T+i{1\over N+1}
    \sum^N_{\beta=1}2^{-\beta+1}\mu_\beta T}
  \sum^\infty_{n^1=-\infty}
  \cdots\sum^\infty_{n^N=-\infty}
  e^{-i{1\over N+1}
    \sum^N_{\beta=1}2^{-\beta+1}2n^\beta\pi}
  \nonumber\\
  &&\times{e^{2i\lambda
      \kakko{2n^\alpha\pi-\mu_\alpha T}}\over
  i\kakko{2n^\alpha\pi-\mu_\alpha T}}
  \prod^N_{\beta=1\atop\beta\ne\alpha}
  {1\over i\left\{2\kakko{n^\alpha-n^\beta}\pi
      -\kakko{\mu_\alpha-\mu_\beta}T\right\}}
  \nonumber\\
  &=&
  e^{-iQc_\alpha T}
  e^{-i{1\over N+1}2^{-N+1}\mu_\alpha T}
  \sum^\infty_{n^\alpha=-\infty}
  {e^{i{1\over N+1}2^{-N+1}2n^\alpha\pi}\over
    i\kakko{2n^\alpha\pi-\mu_\alpha T}}
  \nonumber\\
  &&\times
  \prod^N_{\beta=1\atop\beta\ne\alpha}
  \Bigg[ e^{i{1\over N+1}2^{-\beta+1}\tilde\mu_\beta T}
  \sum^\infty_{\tilde n^\beta=-\infty}
  {e^{i{1\over N+1}2^{-\beta+1}2\tilde n^\beta\pi}\over
    i\kakko{2\tilde n^\beta\pi+\tilde\mu_\beta T}}\Bigg]\ ,
\end{eqnarray}
where $\tilde n^\beta=n^\alpha-n^\beta$ and
$\tilde\mu_\beta\equiv c_\beta-c_\alpha=\mu_\beta-\mu_\alpha,
\kakko{\beta\ne\alpha}$.
We apply the formula
\begin{equation}\label{wakoushiki}
  \sum^\infty_{n=-\infty}
  \frac{e^{2n\pi i\varepsilon}}{2n\pi+\varphi}
  = \frac{e^{i\kakko{\frac{1}{2}-\varepsilon}\varphi}}
  {2\sin\frac{\varphi}{2}}
  \ ,\ \kakko{0<\varepsilon<1}\ ,
\end{equation}
with
\begin{eqnarray}
  &&\varepsilon =
  {1\over N+1}2^{-N+1}\ ,\
  \varphi = -\mu_\alpha T\ \
  {\rm for}\ \alpha\ ,
  \nonumber\\
  &&\varepsilon =
  {1\over N+1}2^{-\beta+1}\ ,\
  \varphi = \tilde\mu_\beta T\ \
  {\rm for}\ \beta\ne\alpha\ ,
\end{eqnarray}
to find
\begin{eqnarray}
  Z_\alpha &=&
  e^{-iQc_\alpha T}
  {1\over1-e^{i\mu_\alpha T}}
  \prod^N_{\beta=1\atop\beta\ne\alpha}
  {1\over1-e^{-i\tilde\mu_\beta T}}
  \nonumber\\
  &=&
  {e^{-iQc_\alpha T}\over\prod^{N+1}_{\beta=1\atop\beta\ne\alpha}
    \kakko{1-e^{-i\tilde\mu_\beta T}}}\ ,
\end{eqnarray}
where $\tilde\mu_{N+1}\equiv c_{N+1}-c_\alpha=-\mu_\alpha$.
Next the calculation of the last term in (\ref{multi:waketa}) is
\begin{eqnarray}
 Z_{N+1} &\equiv&
 e^{-iQc_{N+1}T+i{1\over N+1}
   \sum^N_{\alpha=1} 2^{-\alpha+1}\mu_\alpha T}
 \nonumber\\
 &&\times
 \sum^\infty_{n^1=-\infty}\cdots
 \sum^\infty_{n^N=-\infty}
 e^{-i{1\over N+1}\sum^N_{\alpha=1}
   2^{-\alpha+1}2n^\alpha\pi}
 {\kakko{-1}^N\over\prod^N_{\alpha=1}i
   \kakko{2n^\alpha\pi-\mu_\alpha T}}
 \nonumber\\
 &=&
 e^{-iQc_{N+1}T}\prod^N_{\alpha=1}
 \Bigg[ e^{i{1\over N+1}2^{-\alpha+1}\mu_\alpha T}
 \sum^\infty_{\tilde n^\alpha=-\infty}
 {e^{i{1\over N+1}2^{-\alpha+1}2\tilde n^\alpha\pi}\over
   i\kakko{2\tilde n^\alpha\pi+\mu_\alpha T}}\Bigg]
 \nonumber\\
 &=&
 {e^{-iQc_{N+1}T}\over\prod^N_{\alpha=1}
   \kakko{1-e^{-i\mu_\alpha T}}}\ ,
\end{eqnarray}
where we have put $\tilde n^\alpha=-n^\alpha$ in the
second equality and applied (\ref{wakoushiki}) in the last equality.
Thus
\begin{equation}\label{multi:genmitsu}
  Z = \sum^{N+1}_{\alpha=1}Z_\alpha
    = \sum^{N+1}_{\alpha=1}\frac{e^{-iQc_\alpha T}}
      {\prod^{N+1}_{\beta=1\atop\beta\ne\alpha}
        \left\{1-e^{-i\kakko{c_\beta-c_\alpha}T}\right\}}\ .
\end{equation}
This is the exact result of the trace formula and
just the same with that of the generalized coherent state.
We here note that (\ref{multi:genmitsu}) can be written
in the determinant form:
\begin{eqnarray}\label{multi:genmitsudet}
Z
&=&
\left\vert
\matrix{
  1          & 1          & \cdots & 1\cr
  e^{-ic_1T} & e^{-ic_2T} & \cdots & e^{-ic_{N+1}T}\cr
  \vdots     & \vdots     & \ddots & \vdots\cr
  e^{-i\kakko{N-1}c_1T}   & e^{-i\kakko{N-1}c_2T}
  & \cdots   & e^{-i\kakko{N-1}c_{N+1}T}\cr
  e^{-i\kakko{Q+N}c_1T}   & e^{-i\kakko{Q+N}c_2T}
  & \cdots   & e^{-i\kakko{Q+N}c_{N+1}T}\cr
}\right\vert
\nonumber\\
&&\Bigg/
\left\vert
\matrix{
  1           & 1           & \cdots & 1\cr
  e^{-ic_1T}  & e^{-ic_2T}  & \cdots & e^{-ic_{N+1}T}\cr
  \vdots      & \vdots      & \ddots & \vdots\cr
  e^{-i\kakko{N-1}c_1T}     & e^{-i\kakko{N-1}c_2T}
  & \cdots    & e^{-i\kakko{N-1}c_{N+1}T}\cr
  e^{-iNc_1T} & e^{-iNc_2T} & \cdots & e^{-iNc_{N+1}T}\cr
}\right\vert\ .
\end{eqnarray}
For detail calculation, see appendix \ref{sec:det}.

\section{Relationship to the Generalized Coherent State}
\label{sec:multi:relationship}

In this section we establish a relationship with
(\ref{multi:traceform}) to the trace formula in terms of
the generalized coherent state by an explicit calculation.

By definition, the trace formula in terms of the multi-periodic
coherent state is
\begin{equation}\label{multi:bydef}
  Z =
  \left.\lim_{M\to\infty}\int^{2\pi}_0
    \prod^M_{i=1}d\ffp_i
    \prod^M_{k=1}\bra{\ffp_k}
    \kakko{1-i\delt\hat H}\ket{\ffp_{k-1}}
  \right\vert_{\ffp_M=\ffp_0}\ .
\end{equation}
Inserting the completeness
\begin{equation}
  {\bf 1}_Q =
  \sum_{\| m\|=Q}\ket{m^1,\cdots,m^{N+1}}
  \bra{m^1,\cdots,m^{N+1}}\ ,
\end{equation}
with noting
\begin{eqnarray}
  &&\braket{\ffp}{m^1,\cdots,m^{N+1}} =
  {1\over\kakko{2\pi}^{N/2}}
  e^{i\sum^N_{\alpha=1}m^\alpha\varphi^\alpha}\ ,
  \nonumber\\
  &&\hat H\ket{m^1,\cdots,m^{N+1}} =
  \kakko{\sum^N_{\alpha=1}\mu_\alpha m^\alpha+c_{N+1}Q}
  \ket{m^1,\cdots,m^{N+1}}\ ,
\end{eqnarray}
and putting
\begin{equation}
  \theta_i^\alpha = -\varphi_i^\alpha+2\pi\ ,
\end{equation}
we rewrite (\ref{multi:bydef}) to
\begin{eqnarray}\label{multi:theta}
  Z
  &=&
  \lim_{M\to\infty}\prod^M_{i=1}
  \Bigg[\sum_{\| m_i\|=Q}\int^{2\pi}_0
  {d\fft_i\over\kakko{2\pi}^N}e^{-i\sum^N_{\alpha=1}
    m_i^\alpha\deltheta_i^\alpha}
  \nonumber\\
  &&\left.\times
      \Bigg\{1-i\delt\kakko{\sum^N_{\alpha=1}\mu_\alpha
          m_i^\alpha+c_{N+1}Q}\Bigg\}\Bigg]
  \right\vert_{\varphi_M=\varphi_0}\ ,
\end{eqnarray}
where
\begin{eqnarray}
  \int^{2\pi}_0d\fft_i
  &\equiv&
  \int^{2\pi}_0
  \prod^N_{\alpha=1}
  d\theta^\alpha_i\ ,
  \nonumber\\
  \deltheta_i^\alpha
  &\equiv&
  \theta_i^\alpha-\theta_{i-1}^\alpha\ .
\end{eqnarray}
If we write (\ref{multi:theta}) as
\begin{eqnarray}
  Z
  &=&
  \lim_{M\to\infty}\prod^M_{i=1}
  \Bigg[\sum_{\| m_i\|=Q}\int^{2\pi}_0
    {d\fft_i\over\kakko{2\pi}^N}
    e^{-i\sum^N_{\alpha=1}
      \kakko{m_i^\alpha-m_{i+1}^\alpha}\theta_i^\alpha}
    \nonumber\\
    &&\left.\times
      \Bigg\{1-i\delt\kakko{\sum^N_{\alpha=1}\mu_\alpha
          m_i^\alpha+c_{N+1}Q}\Bigg\}\Bigg]
  \right\vert_{\varphi_M=\varphi_0}\ ,
\end{eqnarray}
we find $m_i^\alpha=m_{i+1}^\alpha$ from the $\fft$-integrals
($\delta$-functions).
By inserting the identity~\cite{FKSF2}
\begin{eqnarray}
  &&{\kakko{N+Q}!\over\ffl!}\int^\infty_0
  \prod^N_{\alpha=1}du^\alpha
  {\kakko{u^1}^{l^1}\cdots\kakko{u^N}^{l^N}\over
    \kakko{1+u^1+\cdots u^N}^{N+Q+1}} = 1\ ,
  \nonumber\\
  &&\ffl! \equiv l^1!\cdots l^{N+1}!\ ,
\end{eqnarray}
into (\ref{multi:theta}), the trace formula becomes
\begin{eqnarray}
Z
&=&
\lim_{M\to\infty}\prod^M_{i=1}\left[\sum_{\| m_i\|=Q}
{\kakko{N+Q}!\over Q!}\int^{2\pi}_0{d\fft_i\over\kakko{2\pi}^N}
\int^\infty_0{\prod^N_{\alpha=1}du_i^\alpha\over
\kakko{1+u_i^1+\cdots+u_i^N}^{N+1}}\right.
\nonumber\\
&&\times
{\kakko{\kakko{u_i^1}^{1/2}e^{-i\theta_i^1}}^{m_i^1}\cdots
\kakko{\kakko{u_i^N}^{1/2}e^{-i\theta_i^N}}^{m_i^N}\over
\kakko{1+u_i^1+\cdots+u_i^N}^{Q/2}}
\nonumber\\
&&\times
{\kakko{\kakko{u_{i-1}^1}^{1/2}e^{i\theta_{i-1}^1}}^{m_i^1}\cdots
\kakko{\kakko{u_{i-1}^N}^{1/2}e^{i\theta_{i-1}^N}}^{m_i^N}\over
\kakko{1+u_{i-1}^1+\cdots+u_{i-1}^N}^{Q/2}}{Q!\over\ffm!}
\nonumber\\
&&\left.\left.\times
\left\{1-i\delt\kakko{\sum^N_{\alpha=1}\mu_\alpha
m_i^\alpha+c_{N+1}Q}\right\}\right]
\right\vert_{\fft_M=\fft_0}\ ,
\end{eqnarray}
where $m_i^\alpha=m_{i+1}^\alpha$ and $u_0^\alpha=u_M^\alpha$
have been used.
Then putting
\begin{equation}
\kakko{u_i^\alpha,\theta_i^\alpha}\longrightarrow\xi_i^\alpha\ ;\
\xi_i^\alpha=\sqrt{u_i^\alpha}e^{i\theta_i^\alpha}\ ,
\end{equation}
we obtain
\begin{eqnarray}
  Z
  &=&
  \lim_{M\to\infty}\prod^M_{i=1}
  \left[\sum_{\| m_i\|=Q}\int
    \measure{\fxi_i}{\fxi_i^\dagger}
    \sqrt{{Q!\over\ffm!}}
    {\kakko{{\xi_i^1}^*}^{m_i^1}\cdots
      \kakko{{\xi_i^N}^*}^{m_i^N}
      \over\kakko{1+\fxi_i^\dagger\fxi_i}^{Q/2}}\right.
  \nonumber\\
  &&\times
  \sqrt{{Q!\over\ffm!}}
  {\kakko{\xi_{i-1}^1}^{m_i^1}\cdots
    \kakko{\xi_{i-1}^N}^{m_i^N}
    \over\kakko{1+\fxi_{i-1}^\dagger\fxi_{i-1}}^{Q/2}}
  \nonumber\\
  &&\left.\left.\times
      \left\{1-i\delt\kakko{\sum^N_{\alpha=1}\mu_\alpha
          m_i^\alpha+c_{N+1}Q}\right\}\right]
  \right\vert_{\fxi_M=\fxi_0}
  \nonumber\\
  &=&
  \lim_{M\to\infty}\prod^M_{i=1}
  \Bigg[\sum_{\| m_i\|=Q}
  \int\measure{\fxi_i}{\fxi_i^\dagger}
  \bra{\fxi_i}\kakko{1-i\delt\hat H}
  \ket{m^1,\cdots,m^{N+1}}
  \nonumber\\
  &&\times
  \braket{m^1,\cdots,m^{N+1}}{\fxi_{i-1}}\Bigg]
  \Bigg\vert_{\fxi_M=\fxi_0}
  \nonumber\\
  &=&
    \lim_{M\to\infty}
    \int\prod^M_{i=1}
    \measure{\fxi_i}{\fxi_i^\dagger}
    \prod^M_{j=1}
    \bra{\fxi_j}\kakko{1-i\delt\hat H}
    \ket{\fxi_{j-1}}
    \bigg\vert_{\fxi_M=\fxi_0}\ ,
\end{eqnarray}
which is nothing but the trace formula in terms of the generalized
coherent state~\cite{FKSF2}.

\section{The WKB Approximation}\label{sec:multi:wkb}

In the Nielsen-Rohrlich formula, the WKB approximation is applicable
to the $\theta$-expression instead of the $p$-expression.
In this section we perform the WKB approximation to the
$\theta$-expression of the trace formula in terms of the
multi-periodic coherent state.
The WKB approximation in this
case is the large $Q\kakko{\lambda}$ expansion.

Writing (\ref{multi:traceform}) to
\begin{eqnarray}
  Z
  &=&
  \sum^\infty_{n^1=-\infty}\cdots
  \sum^\infty_{n^N=-\infty}
  \lim_{M\to\infty}\prod^N_{\alpha=1}\left\{
    \int^{2\kakko{n^\alpha+1}\pi}_{2n^\alpha\pi}
    d\varphi_M^\alpha\right\}
  \int^\infty_{-\infty}\prod^{M-1}_{i=1}d\ffp_i
  \nonumber\\
  &&\times
  \prod^M_{j=1}\left\{\lambda^N\int^\pi_0
    \kakko{\sin^2{\theta_j^1\over2}}^{N-1}
    \sin\theta_j^1{d\theta_j^1\over2\pi}\right.
  \nonumber\\
  &&\left.\times\cdots
    \int^\pi_0
    \kakko{\sin^2{\theta_j^{N-1}\over2}}
    \sin\theta_j^{N-1}{d\theta_j^{N-1}\over2\pi}
    \int^\pi_0\sin\theta_j^N
    {d\theta_j^N\over2\pi}\right\}
  e^{iS^\kakko{n^\alpha}}\ ,
\end{eqnarray}
with the action:
\begin{eqnarray}\label{multi:action}
  S^\kakko{n^\alpha}
  &\equiv&
  -Qc_{N+1}T+{1\over N+1}\sum^N_{\alpha=1}
  2^{-\alpha+1}\mu_\alpha T
  -{1\over N+1}\sum^N_{\alpha=1}
  2^{-\alpha+2}n^\alpha\pi
  \nonumber\\
  &&
  +2\lambda\sum^M_{k=1}\sum^N_{\alpha=1}
  \kakko{\prod^{\alpha-1}_{\beta=1}
    \sin^2{\theta_k^\beta\over2}}
  \cos^2{\theta_k^\alpha\over2}\Phi^\alpha_k\ ,
\end{eqnarray}
we find the equations of motion;
\begin{manyeqns}\label{multi:eqm}
  \bullet&&
  \kakko{\prod^{\alpha-1}_{\beta=1}
    \sin^2{\theta_k^\beta\over2}}
  \cos^2{\theta_k^\alpha\over2}
  -\kakko{\prod^{\alpha-1}_{\beta=1}
    \sin^2{\theta_{k+1}^\beta\over2}}
  \cos^2{\theta_{k+1}^\alpha\over2}=0\ ,
  \label{multi:eqmichi}
  \\
  \bullet&&
  \kakko{\prod^{\alpha-1}_{\beta=1}
    \sin^2{\theta_k^\beta\over2}}
  \sin\theta_k^\alpha\Bigg[-\Phi^\alpha_k
  +\sum^N_{\gamma=\alpha+1}
  \kakko{\prod^{\gamma-1}_{\delta=\alpha+1}
    \sin^2{\theta_k^\delta\over2}}
  \cos^2{\theta_k^\gamma\over2}
  \Phi^\gamma_k\Bigg]=0\ .
  \label{multi:eqmni}
\end{manyeqns}
Now we solve the first equation(s), (\ref{multi:eqmichi}).
In $\alpha=1$ case the equation is
\begin{equation}
  \cos^2{\theta_k^1\over2} =
  \cos^2{\theta_{k+1}^1\over2}\ ,
\end{equation}
and its solution is
\begin{equation}
  \theta_k^1 = \theta_{k+1}^1 =
  \theta^1 = {\rm const}.
\end{equation}
In $\alpha=2$ case the equation is
\begin{equation}
  \sin^2{\theta^1\over2}
  \kakko{\cos^2{\theta_k^2\over2}
    -\cos^2{\theta_{k+1}^2\over2}}=0\ ,
\end{equation}
and its solutions are
\begin{enumerate}
\item $\theta^1=0$, in this case the remaining equations of
  (\ref{multi:eqmichi}) hold with arbitrary $\theta^\alpha$
  for $\alpha\ge2$ ,
\item $\theta^1=C_1\ne0$,
  $\theta_k^2=\theta_{k+1}^2=\theta^2={\rm const}$\ .
\end{enumerate}
By a similar consideration, we obtain the solutions:
\begin{eqnarray}\label{multi:kai1}
  \kakko{\theta^1,\theta^2,\cdots,\theta^N}
  &=&
  \kakko{0,*,*,\cdots,*,*}\ ,
  \nonumber\\
  &=&
  \kakko{C_1,0,*,\cdots,*,*}\ ,
  \nonumber\\
  &=&
  \kakko{C_1,C_2,0,\cdots,*,*}\ ,
  \nonumber\\
  &&\vdots
  \nonumber\\
  &=&
  \kakko{C_1,C_2,C_3,\cdots,C_{N-1},0}\ ,
\end{eqnarray}
where $C_\alpha={\rm constant}\ne0,\kakko{1\le\alpha\le N-1}$ and
$*$ denotes an arbitrary number.
Next we solve (\ref{multi:eqmni}).
The solution $\kakko{C_1,C_2,\cdots,C_{I-1},0,*,\cdots,*}$ of
(\ref{multi:kai1}) satisfies the equations for $\alpha\ge I$
because $\theta_k^I=0$.
In $\alpha=I-1$ case the equation is
\begin{equation}
  \sin C_{I-1}
  \kakko{-\Phi^{I-1}_k+\Phi^I_k} = 0\ .
\end{equation}
Because $-\Phi^{I-1}_k+\Phi^I_k=-\kakko{\delphi_k^{I-1}-\mu_{I-1}\delt}
+\kakko{\delphi_k^I-\mu_I\delt}=0$ is not compatible with
the boundary condition for any given $T$, the solution is
\begin{equation}
  C_{I-1}=\pi\ .
\end{equation}
Therefore we obtain
\begin{equation}
  C_1=C_2=\cdots=C_{I-1}=\pi\ .
\end{equation}
Considering about all $I$'s, finally we obtain the solutions of
(\ref{multi:eqm}):
\begin{eqnarray}
  \kakko{\theta_k^1,\theta_k^2,\cdots,\theta_k^N}
  &=&
  \kakko{0,*,*,\cdots,*,*}\ ,
  \nonumber\\
  &=&
  \kakko{\pi,0,*,\cdots,*,*}\ ,
  \nonumber\\
  &=&
  \kakko{\pi,\pi,0,\cdots,*,*}\ ,
  \nonumber\\
  &&\vdots
  \nonumber\\
  &=&
  \kakko{\pi,\pi,\pi,\cdots,\pi,0}\ ,
  \nonumber\\
  &=&
  \kakko{\pi,\pi,\pi,\cdots,\pi,\pi}\ ,
\end{eqnarray}
with $\varphi$'s being arbitrary for all $\alpha$ and $k$.

First we consider the WKB approximation around
$\kakko{\theta_k^1,\cdots,\theta_k^N}=\kakko{\pi,\pi,\cdots,\pi}$.
Putting
\begin{equation}
  \theta_k^\alpha=\pi-{x_k^\alpha\over\sqrt{\lambda}}\ ,\
  \kakko{\alpha=1,\cdots,N}\ ,
\end{equation}
we write the trace formula to
\begin{eqnarray}
  Z
  &=&
  e^{-iQc_{N+1}T+i{1\over N+1}\sum^N_{\alpha=1}
    2^{-\alpha+1}\mu_\alpha T}
  \nonumber\\
  &&\times
  \sum^\infty_{n^1=-\infty}\cdots
  \sum^\infty_{n^N=-\infty}
  e^{-i{1\over N+1}\sum^N_{\alpha=1}
    2^{-\alpha+2}n^\alpha\pi}
  \nonumber\\
  &&\times
  \lim_{M\to\infty}\prod^N_{\alpha=1}
  \left\{\int^{2\kakko{n^\alpha+1}\pi}
    _{2n^\alpha\pi}d\varphi_M^\alpha\right\}
  \int^\infty_{-\infty}\prod^{M-1}_{i=1}d\ffp_i
  \nonumber\\
  &&\times
  \prod^M_{j=1}\left\{\lambda^N\int^\infty_0
    \kakko{\cos^2{x_j^1\over2\sqrt{\lambda}}}^{N-1}
      \sin{x_j^1\over\sqrt{\lambda}}
    {dx_j^1\over2\pi\sqrt{\lambda}}\cdots\right.
  \nonumber\\
  &&\left.\times
    \int^\infty_0\cos^2{x_j^{N-1}\over2\sqrt{\lambda}}
      \sin{x_j^{N-1}\over\sqrt{\lambda}}
      {dx_j^{N-1}\over2\pi\sqrt{\lambda}}
    \int^\infty_0\sin{x_j^N\over\sqrt{\lambda}}
    {dx_j^N\over2\pi\sqrt{\lambda}}\right\}
  \nonumber\\
  &&\times
  \exp\left[2i\lambda\sum^M_{k=1}\sum^N_{\alpha=1}
    \kakko{\prod^{\alpha-1}_{\beta=1}
      \cos^2{x_k^\beta\over2\sqrt{\lambda}}}
    \kakko{\sin^2{x_k^\alpha\over2\sqrt{\lambda}}}
    \Phi^\alpha_k\right]\ .
\end{eqnarray}
The leading order term $\kakko{\kansu{O}{(1/\lambda)^0}}$ becomes
\begin{eqnarray}\label{multi:leadmoto}
  Z_{N+1}
  &\equiv&
  e^{-iQc_{N+1}T+i{1\over N+1}\sum^N_{\alpha=1}
    2^{-\alpha+1}\mu_\alpha T}
  \nonumber\\
  &&\times
  \sum^\infty_{n^1=-\infty}\cdots
  \sum^\infty_{n^N=-\infty}
  e^{-i{1\over N+1}\sum^N_{\alpha=1}
    2^{-\alpha+2}n^\alpha\pi}
  \nonumber\\
  &&\times
  \lim_{M\to\infty}\prod^N_{\alpha=1}
  \left[\int^{2\kakko{n^\alpha+1}\pi}_{2n^\alpha\pi}
    {d\varphi_M^\alpha\over2\pi}
    \int^\infty_{-\infty}\prod^{M-1}_{i=1}
    {d\varphi_i^\alpha\over2\pi}
    \int^\infty_0\prod^M_{j=1}x_j^\alpha dx_j^\alpha\right.
  \nonumber\\
  &&\left.\times
    \exp\left\{{i\over2}\sum^M_{k=1}
      \kakko{x_k^\alpha}^2
      \kakko{\Phi^\alpha_k+i\delta_M}\right\}\right]\ ,
\end{eqnarray}
where $\delta_M$ has been introduced, which is given for each $M$ and
finally put zero, to ensure the convergence of the $x$-integrals.
By the form of (\ref{multi:leadmoto}), each $x_j^\alpha$-integral
can be performed independently to give
\begin{eqnarray}
  Z_{N+1}
  &=&
  e^{-iQc_{N+1}T+i{1\over N+1}
    \sum^N_{\alpha=1}2^{-\alpha+1}\mu_\alpha T}
  \sum^\infty_{n^1=-\infty}\cdots
  \sum^\infty_{n^N=-\infty}
  e^{-i{1\over N+1}\sum^N_{\alpha=1}
    2^{-\alpha+2}n^\alpha\pi}
  \nonumber\\
  &&\times
  \lim_{M\to\infty}\prod^N_{\alpha=1}
  \left[\int^{2\kakko{n^\alpha+1}\pi}_{2n^\alpha\pi}
    {d\varphi_M^\alpha\over2\pi}
    \int^\infty_{-\infty}\prod^{M-1}_{i=1}
    {d\varphi^\alpha_i\over2\pi}\prod^M_{j=1}
    {1\over i\kakko{\mu_\alpha\delt
        -\delphi_j^\alpha-i\delta_M}}\right]\ .
\end{eqnarray}
Noting that
\begin{eqnarray}\label{phiseibun}
  &&\int^\infty_{-\infty}
  {d\varphi^\alpha_i\over2\pi i}
  {1\over\mu_\alpha\delt-\varphi_{j+1}^\alpha
    +\varphi_j^\alpha-i\delta_M}
  {1\over\mu_\alpha\delt-\varphi_j^\alpha
    +\varphi_{j-1}^\alpha-i\delta_M}
  \nonumber\\
  &&=
  {1\over\mu_\alpha2\delt-\varphi_{j+1}^\alpha
    +\varphi_{j-1}^\alpha-2i\delta_M}\ ,
\end{eqnarray}
we perform the $\varphi$-integrals to obtain
\begin{eqnarray}
  Z_{N+1}
  &=&
  e^{-iQc_{N+1}T+i{1\over N+1}
    \sum^N_{\alpha=1}2^{-\alpha+1}\mu_\alpha T}
  \prod^N_{\alpha=1}\left[
    i\sum^\infty_{n^\alpha=-\infty}
    {e^{i{1\over N+1}2^{-\alpha+2}n^\alpha\pi}
      \over2n^\alpha\pi-\mu_\alpha T}\right]
  \nonumber\\
  &=&
  e^{-iQc_{N+1}T}
  \prod^N_{\alpha=1}\left[
    e^{i{1\over N+1}2^{-\alpha+1}\mu_\alpha T}
    \sum^\infty_{\tilde n^\alpha=-\infty}
    {e^{i2\tilde n^\alpha\pi{1\over N+1}2^{-\alpha+1}}
      \over i\kakko{2\tilde n^\alpha\pi+\mu_\alpha T}}\right]\ ,
\end{eqnarray}
where we have put $\delta_M=0$ and $\tilde n^\alpha=-n^\alpha$
in the second equality.
Application of the formula  (\ref{wakoushiki}) with noting
\begin{equation}
  0<{1\over N+1}2^{-\alpha+1}<1\ ,
\end{equation}
gives the final form
\begin{eqnarray}
  Z_{N+1}
  &=&
  e^{-iQc_{N+1}T}
  \prod^N_{\alpha=1}\left[
    e^{i{1\over N+1}2^{-\alpha+1}\mu_\alpha T}
    {e^{i\kakko{{1\over2}-{2^{-\alpha+1}\over N+1}}
        \mu_\alpha T}\over
      2i\sin{\mu_\alpha T\over2}}\right]
  \nonumber\\
  &=&
  {e^{-iQc_{N+1}T}\over\prod^N_{\alpha=1}
    \kakko{1-e^{-i\mu_\alpha T}}}
  \nonumber\\
  &=&
  {e^{-iQc_{N+1}T}\over
    \prod^N_{\alpha=1}\left\{1
      -e^{-i\kakko{c_\alpha-c_{N+1}}T}\right\}}\ .
\end{eqnarray}

Next we perform the WKB approximation around
\begin{equation}\label{kaii}
 \kakko{\theta^1_k,\cdots,\theta^N_k} =
 (\stackrel{1}{\stackrel{ }{\pi}},\stackrel{2}{\stackrel{ }{\pi}},
   \cdots,\stackrel{I-1}{\stackrel{ }{\pi}},\stackrel{I}{\stackrel{ }{0}},
   \stackrel{I+1}{\stackrel{ }{*}},\cdots,\stackrel{N}{\stackrel{ }{*}})\ ,
\end{equation}
for $I=1,\cdots,N$. ($(\theta^1_k,\cdots,\theta^N_k)=
(\pi,\pi,\cdots,\pi,0)$ is a special case of
(\ref{kaii}).)
We put
\begin{equation}
  \theta^\alpha_k = \pi-
  {x^\alpha_k\over\sqrt{\lambda}}\ ,\
  \kakko{\alpha=1,\cdots,I-1}\ ,\
  \theta^I_k={x^I_k\over\sqrt{\lambda}}\ ,
\end{equation}
and leave $\theta^\alpha_k(\alpha=I+1,\cdots,N)$ unchanged
because they are arbitrary numbers to have no expansion points.
The leading order term of the trace formula then becomes
\begin{eqnarray}\label{pakari}
  Z_I &\equiv&
  e^{-iQc_{N+1}T+i{1\over N+1}
    \sum^N_{\alpha=1}2^{-\alpha+1}\mu_\alpha T}
  \sum^\infty_{n^1=-\infty}
  \cdots\sum^\infty_{n^N=-\infty}
  e^{-i{1\over N+1}\sum^N_{\alpha=1}
    2^{-\alpha+1}2n^\alpha\pi}
  \nonumber\\
  &&\times
  \lim_{M\to\infty}\prod^N_{\alpha=1}\bigg\{
  \int^{2\kakko{n^\alpha+1}\pi}_{2n^\alpha\pi}
  d\varphi^\alpha_M\bigg\}
  \int^\infty_{-\infty}\prod^{M-1}_{i=1}d\ffp_i
  \nonumber\\
  &&\times
  \prod^M_{j=1}\bigg\{\int^\infty_0x^1_j
  {dx^1_j\over2\pi}\cdots
  \int^\infty_0x^{I-1}_j{dx^{I-1}_j\over2\pi}
  \int^\infty_02^{-2\kakko{N-I}}
  \kakko{x^I_j}^{2\kakko{N-I}+1}
  {dx^I_j\over2\pi}
  \nonumber\\
  &&\times
  \int^\pi_0\kakko{\sin^2{\theta^{I+1}_j\over2}}^{N-I-1}
  \sin\theta^{I+1}_j{d\theta^{I+1}_j\over2\pi}\cdots
  \int^\pi_0\sin\theta^N_j{d\theta^N_j\over2\pi}\bigg\}
  e^{2i\lambda\kakko{2n^I\pi-\mu_IT}}
  \nonumber\\
  &&\times
  \exp\bigg[{i\over2}\sum^M_{k=1}\sum^{I-1}_{\alpha=1}
  \kakko{x^\alpha_k}^2\kakko{\Phi^\alpha_k-\Phi^I_k}
  -{i\over2}\sum^M_{k=1}\kakko{x^I_k}^2\Phi^I_k
  \nonumber\\
  &&
  +{i\over2}\sum^M_{k=1}
  \kakko{x^I_k}^2\sum^N_{\alpha=I+1}
  \kakko{\prod^{\alpha-1}_{\beta=I+1}
    \sin^2{\theta^\beta_k\over2}}
  \cos^2{\theta^\alpha_k\over2}\Phi^\alpha_k\bigg]
  \nonumber\\
  &=&
  e^{-iQc_IT}\hat Z_Ie^{-i{1\over N+1}
    2^{-I+1}\mu_IT+i{1\over N+1}
    \sum^N_{\alpha=I+1}2^{-\alpha+1}\mu_\alpha T}
  \nonumber\\
  &&\times
  \sum^\infty_{n^I=-\infty}\cdots
  \sum^\infty_{n^N=-\infty}
  e^{i{1\over N+1}2^{-I+1}2n^I\pi
    -{1\over N+1}\sum^N_{\alpha=I+1}
    2^{-\alpha+1}2n^\alpha\pi}
  \nonumber\\
  &&\times
  \lim_{M\to\infty}\prod^N_{\alpha=I}
  \bigg\{\int^{2\kakko{n^\alpha+1}\pi}_{2n^\alpha\pi}
  {d\varphi^\alpha_M\over2\pi}\int\prod^{M-1}_{i=1}
  {d\varphi^\alpha_i\over2\pi}\bigg\}
  \nonumber\\
  &&\times
  \prod^M_{j=1}\bigg\{2^{-\kakko{N-I}}\int^\infty_0
  \kakko{x^I_j}^{2\kakko{N-I}+1}dx^I_j
  e^{-{i\over2}\kakko{x^I_j}^2\Phi^I_j}
  \nonumber\\
  &&\times
  \int^\pi_0\kakko{\sin^2{\theta^{I+1}_j\over2}}^{N-I-1}
  \sin\theta^{I+1}_jd\theta^{I+1}_j\cdots
  \int^\pi_0\sin\theta^N_jd\theta^N_j
  \nonumber\\
  &&\times
  \exp\Big[{i\over2}\kakko{x^I_j}^2
  \sum^N_{\alpha=I+1}
  \kakko{\prod^{\alpha-1}_{\beta=I+1}
    \sin^2{\theta^\beta_j\over2}}
  \cos^2{\theta^\alpha_j\over2}\Phi^\alpha_j
  \Big]\bigg\}\ ,
\end{eqnarray}
where
\begin{eqnarray}
  \hat Z_I &\equiv&
  e^{i{1\over N+1}\sum^{I-1}_{\alpha=1}
    2^{-\alpha+1}\tilde\mu_\alpha T}
  \sum^\infty_{n^1=-\infty}\cdots
  \sum^\infty_{n^{I-1}=-\infty}
  e^{i{1\over N+1}\sum^{I-1}_{\alpha=1}
    2^{-\alpha+1}2\kakko{n^I-n^\alpha}\pi}
  \nonumber\\
  &&\times
  \lim_{M\to\infty}\prod^{I-1}_{\alpha=1}
  \bigg\{\int^{2\kakko{n^\alpha+1}\pi}_{2n^\alpha\pi}
  {d\varphi^\alpha_M\over2\pi}
  \int^\infty_{-\infty}\prod^{M-1}_{i=1}
  {d\varphi^\alpha\over2\pi}
  \int^\infty_0\prod^M_{j=1}x^\alpha_jdx^\alpha_j
  \nonumber\\
  &&\times
  \exp\bigg[{i\over2}\sum^M_{k=1}
  \kakko{x^\alpha_k}^2
  \kakko{\Phi^\alpha_k-\Phi^I_k}\bigg]\Bigg\}\ ,
\end{eqnarray}
with $\tilde\mu_\alpha\equiv\mu_\alpha-\mu_I$.
We can calculate $\hat Z_I$ by the similar way with
$\kakko{\theta^1_k,\cdots,\theta^N_k}=\kakko{\pi,\cdots,\pi}$
case by substituting $N\to I-1$ and
$\Phi^\alpha_k\to\Phi^\alpha_k-\Phi^I_k$ to obtain
\begin{eqnarray}\label{zigutai}
  \hat Z_I &=&
  e^{i{1\over N+1}\sum^{N-1}_{\alpha=1}
    2^{-\alpha+1}\tilde\mu_\alpha T}
  \sum^\infty_{n^1=-\infty}\cdots
  \sum^\infty_{n^{I-1}=-\infty}
  e^{i{1\over N+1}\sum^{I-1}_{\alpha=1}
    2^{-\alpha+1}2\kakko{n^I-n^\alpha}\pi}
  \nonumber\\
  &&\times
  \prod^{I-1}_{\alpha=1}\Bigg[
  {1\over-i\left\{2\kakko{n^\alpha-n^I}\pi
      -\kakko{\mu_\alpha-\mu_I}T\right\}}\Bigg]
  \nonumber\\
  &=&
  \prod^{I-1}_{\alpha=1}
  \Bigg[ e^{i{1\over N+1}
    2^{-\alpha+1}\tilde\mu_\alpha T}
  \sum^\infty_{\tilde n^\alpha=-\infty}
  {e^{i{1\over N+1}2^{-\alpha+1}2\tilde n^\alpha\pi}\over
  i\kakko{2\tilde n^\alpha+\tilde\mu_\alpha T}}\Bigg]
  \nonumber\\
  &=&
  \prod^{I-1}_{\alpha=1}{1\over 1-e^{-i\tilde\mu_\alpha T}}\ ,
\end{eqnarray}
where $\tilde n^\alpha=n^I-n^\alpha$.
For the remaining part of (\ref{pakari}),
changing variables such that
\begin{equation}
  p^\alpha_k =
  \kakko{\prod^{\alpha-1}_{\beta=I+1}
    \sin^2{\theta^\beta_k\over2}}
  \cos^2{\theta^\alpha_k\over2}\ ,\
  \alpha=I+1,\cdots,N\ ,
\end{equation}
with the Jacobian
\begin{equation}
  {\kansu{\partial}{p}\over
    \kansu{\partial}{\theta}} =
  \kakko{-1}^{N-I}2^{-N+I}
  \prod^N_{\alpha=I+1}\Bigg\{
  \kakko{\sin^2{\theta^\alpha_k\over2}}^{N-\alpha}
  \sin\theta^\alpha_k\Bigg\}\ ,
\end{equation}
gives the form of the leading order term
\begin{eqnarray}\label{multi:umae}
  Z_I &=&
  e^{-iQc_IT}\hat Z_I
  e^{-i{1\over N+1}2^{-I+1}\mu_IT
    +i{1\over N+1}\sum^N_{\alpha=I+1}
    2^{-\alpha+1}\mu_\alpha T}
  \nonumber\\
  &&\times
  \sum^\infty_{n^I=-\infty}\cdots
  \sum^\infty_{n^N=-\infty}
  e^{i{1\over N+1}2^{-I+1}2n^I\pi
    -i{1\over N+1}\sum^N_{\alpha=I+1}
    2^{-\alpha+1}2n^\alpha\pi}
  \nonumber\\
  &&\times
  \lim_{M\to\infty}\prod^N_{\alpha=1}\bigg\{
  \int^{2\kakko{n^\alpha+1}\pi}_{2n^\alpha\pi}
  \frac{d\varphi^\alpha_M}{2\pi}
  \int^\infty_{-\infty}\prod^{M-1}_{i=1}
  \frac{d\varphi^\alpha_i}{2\pi}
  \bigg\}
  \nonumber\\
  &&\times
  \prod^M_{j=1}\bigg\{\int^\infty_0
  \kakko{x^I_j}^{2\kakko{N-I}+1}
  dx^I_je^{-\frac{i}{2}\kakko{x^I_j}^2\Phi^I_j}
  \nonumber\\
  &&\times
  \int^1_0dp^{I+1}_j\int^{1-p^{I+1}_j}_0dp^{I+2}_j\cdots
  \int^{1-\sum^{N-1}_{\alpha=I+1}p^\alpha_j}_0dp^N_j
  \nonumber\\
  &&\times
  \exp\Big[\frac{i}{2}\kakko{x^I_j}^2\sum^N_{\alpha=I+1}
  p^\alpha_j\Phi^\alpha_j\Big]\bigg\}\ .
\end{eqnarray}
By putting $u_j={1\over2}(x^I_j)^2$ and applying the integral
relation (\ref{sekibunkoushiki}) with $L=I+1,u=u_j,
\Phi^\alpha=\Phi^\alpha_j$, $u$-integrals becomes
\begin{eqnarray}
  &&\int^\infty_0\kakko{u_j}^{N-I}du_je^{-iu_j\Phi^I_j}
  \int^1_0dp^{I+1}_j\cdots
  \int^{1-\sum^{N-1}_{\alpha=I+1}p^\alpha_j}_0dp^N_j
  e^{iu_j\sum^N_{\alpha=I+1}p^\alpha_j\Phi^\alpha_j}
  \nonumber\\
  &=&
  \int^\infty_0\kakko{u_j}^{N-I}du_je^{-iu_j\Phi^I_j}
  \kakko{iu_j}^{-N+\kakko{I+1}-1}
  \nonumber\\
  &&\times
  \bigg[\sum^N_{\alpha=I+1}
  {1\over\Phi^\alpha_j\prod^N_{\beta=I+1\atop\beta\ne\alpha}
    \kakko{\Phi^\alpha_j-\Phi^\beta_j}}e^{iu_j\Phi^\alpha_j}
  +{\kakko{-1}^{N-\kakko{I-1}+1}\over
  \prod^N_{\beta=I+1}\Phi^\beta_j}\bigg]
  \nonumber\\
  &=&
  \lim_{\delta\to0}
  i^{-N+I}\Bigg[\sum^N_{\alpha=I+1}
  {1\over\Phi^\alpha_j\prod^N_{\beta=I+1\atop\beta\ne\alpha}
  \kakko{\Phi^\alpha_j-\Phi^\beta_j}}
  \int^\infty_0du_je^{iu_j\kakko{\Phi^\alpha_j-\Phi^I_j+i\delta}}
  \nonumber\\
  &&+
  {\kakko{-1}^{N-I}\over\prod^N_{\beta=I+1}\Phi^\beta_j}
  \int^\infty_0du_je^{-iu_j\kakko{\Phi^I_j-i\delta}}\Bigg]
  \nonumber\\
  &=&
  i^{-N+I-1}\bigg\{-\sum^N_{\alpha=I+1}
  {1\over\Phi^\alpha_j\prod^N_{\beta=I\atop\beta\ne\alpha}
    \kakko{\Phi^\alpha_j-\Phi^\beta_j}}
  +{\kakko{-1}^{N-I}\over\prod^N_{\beta=I}\Phi^\beta_j}\bigg\}
  \nonumber\\
  &=&
  {1\over i\Phi^I_j\prod^N_{\beta=I+1}i
    \kakko{\Phi^I_j-\Phi^\beta_j}}\ ,
\end{eqnarray}
where the regularization parameter $\delta$ has been introduced
for the $u_j$-integrals to converge in the third equality and
the relation (\ref{dethenksei}) has been applied
in the last equality.
Thus (\ref{multi:umae}) becomes
\begin{eqnarray}
  Z_I &=&
  e^{-iQc_IT}\hat Z_I
  e^{-i{1\over N+1}2^{-I+1}\mu_IT
    +i{1\over N+1}\sum^N_{\alpha=I+1}
    2^{-\alpha+1}\mu_\alpha T}
  \nonumber\\
  &&\times
  \sum^\infty_{n^I=-\infty}\cdots
  \sum^\infty_{n^N=-\infty}
  e^{i{1\over N+1}2^{-I+1}2n^I\pi
    -i{1\over N+1}\sum^N_{\alpha=I+1}
    2^{-\alpha+1}2n^\alpha\pi}
  \nonumber\\
  &&\times
  \lim_{M\to\infty}
  \int^{2\kakko{n^I+1}\pi}_{2n^I\pi}
  {d\varphi^I_M\over2\pi}
  \int^\infty_{-\infty}\prod^{M-1}_{i=1}
  {d\varphi^I_i\over2\pi}
  \prod^M_{j=1}{1\over i\Phi^I_j}
  \nonumber\\
  &&\times
  \prod^N_{\alpha=I+1}\left\{
  \int^{2\kakko{n^\alpha+1}\pi}_{2n^\alpha\pi}
  {d\varphi^\alpha_M\over2\pi}
  \int^\infty_{-\infty}\prod^{M-1}_{i=1}
  {d\varphi^\alpha_i\over2\pi}
  \prod^M_{j=1}
  {1\over i\kakko{\Phi^I_j-\Phi^\alpha_j}}\right\}
  \nonumber\\
  &=&
  e^{-iQc_IT}\hat Z_I
  e^{-i{1\over N+1}2^{-I+1}\mu_IT
    +i{1\over N+1}\sum^N_{\alpha=I+1}
    2^{-\alpha+1}\mu_\alpha T}
  \nonumber\\
  &&\times
  \sum^\infty_{n^I=-\infty}\cdots
  \sum^\infty_{n^N=-\infty}
  e^{i{1\over N+1}2^{-I+1}2n^I\pi
    -i{1\over N+1}\sum^N_{\alpha=I+1}
    2^{-\alpha+1}2n^\alpha\pi}
  \nonumber\\
  &&\times
  {1\over i\kakko{2n^I\pi-\mu_IT}}
  \prod^N_{\alpha=I+1}
  {1\over i\left\{
      2\kakko{n^I-n^\alpha}\pi
      -\kakko{\mu_I-\mu_\alpha}T\right\}}\ ,
\end{eqnarray}
where the $\varphi$-integrals have been performed
by noting (\ref{phiseibun}).
Putting
\begin{eqnarray}
  &&\tilde n^\alpha = n^I-n^\alpha\ ,\
  \tilde\mu_\alpha\equiv c_\alpha-c_I =
  \mu_\alpha-\mu_I\ ,\
  \kakko{\alpha=I+1,\cdots,N}\ ,
  \nonumber\\
  &&\tilde\mu_{N+1}\equiv c_{N+1}-c_I
  = -\mu_I\ ,
\end{eqnarray}
we obtain
\begin{eqnarray}
  Z_I &=&
  e^{-iQc_IT}\hat Z_I
  e^{i{1\over N+1}2^{-N+1}\tilde\mu_{N+1}T}
  \sum^\infty_{n^I=-\infty}
  {e^{i{1\over N+1}2^{-N+1}2n^I\pi}\over
    i\kakko{2n^I\pi+\tilde\mu_{N+1}T}}
  \nonumber\\
  &&\times
  \prod^N_{\alpha=I+1}\Bigg[
  e^{i{1\over N+1}2^{-\alpha+1}\tilde\mu_\alpha T}
  \sum^\infty_{\tilde n^\alpha=-\infty}
  {e^{i{1\over N+1}2^{-\alpha+1}2\tilde n^\alpha\pi}
    \over i\kakko{2\tilde n^\alpha\pi
      +\tilde\mu_\alpha T}}\Bigg]
  \nonumber\\
  &=&
  e^{-iQc_IT}\hat Z_I
  e^{i{1\over N+1}2^{-N+1}\tilde\mu_{N+1}T}
  {e^{i\kakko{{1\over2}-{1\over N+1}2^{-N+1}}
      \tilde\mu_{N+1}T}\over
    2i\sin{\tilde\mu_{N+1}T\over2}}
  \nonumber\\
  &&\times
  \prod^N_{\alpha=I+1}\Bigg[
  e^{i{1\over N+1}2^{-\alpha+1}\tilde\mu_\alpha T}
  {e^{i\kakko{{1\over2}-{1\over N+1}2^{-\alpha+1}}
        \tilde\mu_\alpha T}\over
      2i\sin{\tilde\mu_\alpha T\over2}}\Bigg]
    \nonumber\\
    &=&
    e^{-iQc_IT}\hat Z_I
    {1\over1-e^{-i\tilde\mu_{N+1}T}}
    \prod^N_{\alpha=I+1}
    {1\over1-e^{-i\tilde\mu_\alpha T}}
    \nonumber\\
    &=&
    {e^{-iQc_IT}\over
      \prod^{N+1}_{\alpha=1\atop\alpha\ne I}
      \kakko{1-e^{-i\tilde\mu_\alpha T}}}\ ,
\end{eqnarray}
where we have applied the formula (\ref{wakoushiki})
in the second equality and put the explicit form of
$\hat Z_I$, (\ref{zigutai}), in the last equality.
Thus the total contribution of the WKB approximation becomes
\begin{equation}
  Z_{\rm WKB} =
  \sum^{N+1}_{\alpha=1}Z_\alpha =
  \sum^{N+1}_{\alpha=1}
  {e^{-iQc_\alpha T}\over
    \prod^{N+1}_{\beta=1\atop\beta\ne\alpha}
    \left\{1-e^{-i\kakko{c_\beta-c_\alpha}T}\right\}}\ ,
\end{equation}
which is the same with the exact calculation (\ref{multi:genmitsu}).
Therefore we can conclude that {\em the WKB-exactness holds in the
multi-periodic coherent state}.

\section{Discussion}\label{sec:discussion}

In this paper we have constructed the trace formula in terms of the
multi-periodic coherent state as an extension of the
Nielsen-Rohrlich formula for spin.
We have made an exact calculation of the trace formula
and performed the WKB approximation to show the WKB-exactness.
We have also clarified a connection between the trace formula
and that of the generalized coherent state.

The result obtained in this paper is perfectly parallel with
the generalized coherent state~\cite{FKSF2}.
Similar argument in terms of another coherent state may be
possible.
However the WKB-exactness will not hold in arbitrary
coherent states.
Now we have the problem that {\em what kind of coherent states
make the system WKB-exact.}

The extension to the Grassmann manifold from $CP^N$
have been made~\cite{FKS}.
The corresponding extension from the multi-periodic coherent state
will be also possible.

The extension to supersymmetric $CP^1(CP^N)$ model is also attractive.
There is an expectation that by the fermion contribution
the result of the continuum path integral coincides with
the discrete one.

The extension to field theories is more important.
There are some attempts in this field~\cite{Witten-Yang,Perret,BlauThom}.
However they seem to be still insufficient in the mathematical
point of view.

\begin{center}
  {\bf Acknowledgments}
\end{center}
We wish to thank T. Kashiwa and S. Sakoda for useful discussions.

\newpage
\begin{center}
{\Large\bf Appendix}
\end{center}

\appendix

\section{Some Useful Relations}\label{sec:koshiki}

\subsection{The Vandermonde's Determinant}

We define
\begin{eqnarray}
  \Delta_k &\equiv&
  \prod^n_{\alpha=k}\prod^{\alpha-1}_{\beta=k}
  \kakko{a_\alpha-a_\beta} =
  \left\vert\matrix{1         & 1             & \cdots & 1\cr
                    a_k       & a_{k+1}       & \cdots & a_n\cr
                    a_k^2     & a_{k+1}^2     & \cdots & a_n^2\cr
                    \vdots    & \vdots        & \ddots & \vdots\cr
                    a_k^{n-k} & a_{k+1}^{n-k} & \cdots & a_n^{n-k}\cr}
  \right\vert\ ,
  \nonumber\\
  \kansu{\Delta_k}{\alpha} &\equiv&
  \left\vert\matrix{
      a_k                & a_{k+1}           & \cdots &
      a_{\alpha-1}       & a_{\alpha+1}      & \cdots & a_n\cr
      a_k^2              & a_{k+1}^2         & \cdots &
      a_{\alpha-1}^2     & a_{\alpha+1}^2    & \cdots & a_n^2\cr
      \vdots             & \vdots            & \ddots &
      \vdots             & \vdots            & \ddots & \vdots\cr
      a_k^{n-k}          & a_{k+1}^{n-k}     & \cdots &
      a_{\alpha-1}^{n-k} & a_{\alpha+1}^{n-k} & \cdots & a_n^{n-k}\cr}
  \right\vert\ ,
  \nonumber\\
  \kansu{\tilde\Delta_k}{\alpha} &\equiv&
  \left\vert\matrix{1 & \cdots & 1 & 1 & \cdots & 1\cr
                    a_k & \cdots & a_{\alpha-1} & a_{\alpha+1}
                    & \cdots & a_n\cr
                    a_k^2 & \cdots & a_{\alpha-1}^2 & a_{\alpha+1}^2
                    & \cdots & a_n^2\cr
                    \vdots & & \vdots & \vdots & & \vdots\cr
                    a_k^{n-k-1} & \cdots &a_{\alpha-1}^{n-k-1}
                    & a_{\alpha+1}^{n-k-1}&\cdots&a_n^{n-k-1}\cr}
  \right\vert
  =
  \prod^n_{\beta=k\atop\beta\ne\alpha}
  \prod^{\beta-1}_{\gamma=k\atop\gamma\ne\alpha}
  \kakko{a_\beta-a_\gamma}\ ,
  \nonumber\\
  P_k &\equiv& \prod^n_{\alpha=k}a_\alpha = a_ka_{k+1}\cdots a_n\ .
\end{eqnarray}
$\Delta_k$ is known as the Vandermonde's determinant
and $\kansu{\Delta_k}{\alpha}$ is its cofactor
apart from sign factor $(-1)^{\alpha-k}$.
We then find the relations (Laplace expansion):
\begin{eqnarray}\label{multi:yoinshi}
  &&\Delta_k =
  \sum^n_{\alpha=k}\kakko{-1}^{\alpha-k}
  \kansu{\Delta_k}{\alpha}\ ,
  \nonumber\\
  &&\kansu{\Delta_k}{\alpha} =
  {P_k\over a_\alpha}
  \kansu{\tilde\Delta_k}{\alpha}\ .
\end{eqnarray}
We rewrite $\kansu{\Delta_k}{\alpha}$ by another expression.
Since
\begin{equation}\label{app:youso}
  \prod^n_{\beta=k\atop\beta\ne\alpha}
  \kakko{a_\alpha-a_\beta}
  =
  \kakko{-1}^{n-\alpha}
  \Delta_k{1\over\kansu{\tilde\Delta_k}{\alpha}}
  =
  \kakko{-1}^{n-\alpha}{P_k\over a_\alpha}
  {\Delta_k\over\kansu{\Delta_k}{\alpha}}\ ,
\end{equation}
we have
\begin{equation}\label{gyakutoki}
  \kansu{\Delta_k}{\alpha} =
  \kakko{-1}^{n-\alpha}
  {P_k\over a_\alpha
    \prod^n_{\beta=k\atop\beta\ne\alpha}
    \kakko{a_\alpha-a_\beta}}
  \Delta_k\ .
\end{equation}
Substituting (\ref{gyakutoki}) into the first relation in
(\ref{multi:yoinshi}), we then find
\begin{equation}
  \sum^n_{\alpha=k}{P_k\over a_\alpha
    \prod^n_{\beta=k\atop\beta\ne\alpha}
    \kakko{a_\alpha-a_\beta}}
  = \kakko{-1}^{n-k}\ ,
\end{equation}
and, picking out $\alpha=k$ term,
\begin{equation}\label{dethenksei}
  {1\over a_k\prod^n_{\beta=k+1}\kakko{a_k-a_\beta}} =
  -\sum^n_{\alpha=k+1}{1\over a_\alpha\prod^n_{\beta=k\atop\beta\ne\alpha}
      \kakko{a_\alpha-a_\beta}}
    +{\kakko{-1}^{n-k}\over P_k}\ .
\end{equation}

\subsection{Some Integral}

We claim
\begin{eqnarray}\label{sekibunkoushiki}
  \kansu{A}{L} &\equiv&
  \int^{1-\sum^{L-1}_{\gamma=I+1}p^\gamma}_0
  dp^Le^{iu\Phi^Lp^L}
  \cdots
  \int^{1-\sum^{N-1}_{\gamma=I+1}p^\gamma}_0
  dp^Ne^{iu\Phi^Np^N}
  \nonumber\\
  &=&
  \kakko{iu}^{-N+L-1}\bigg[\sum^N_{\alpha=L}
  {1\over\Phi^\alpha\prod^N_{\beta=L\atop\beta\ne\alpha}
    \kakko{\Phi^\alpha-\Phi^\beta}}
  e^{iu\kakko{1-\sum^{L-1}_{\gamma=I+1}p^\gamma}\Phi^\alpha}
  +{\kakko{-1}^{N-L+1}\over\prod^N_{\beta=L}\Phi^\beta}\bigg]\ ,
  \nonumber\\
  &&{\rm for}\ L=N,N-1,\cdots,I+1\ .
\end{eqnarray}
We prove this relation by mathematical induction.
In $L=N$ case we can examine (\ref{sekibunkoushiki}) by an explicit
calculation.
Now we assume that (\ref{sekibunkoushiki}) holds in $L$.
We then calculate in $L-1$ case:
\begin{eqnarray}\label{tsugi}
  \kansu{A}{L-1} &\equiv&
  \int^{1-\sum^{L-2}_{\gamma=I+1}p^\gamma}_0
  dp^{L-1}e^{iu\Phi^{L-1}p^{L-1}}
  \cdots
  \int^{1-\sum^{N-1}_{\gamma=I+1}p^\gamma}_0
  dp^Ne^{iu\Phi^Np^N}
  \nonumber\\
  &=&
  \int^{1-\sum^{L-2}_{\gamma=I+1}p^\gamma}_0
  dp^{L-1}\kansu{A}{L}
  \nonumber\\
  &=&
  \kakko{-1}^{-N+L-1}\bigg[\sum^N_{\alpha=L}
  {1\over\Phi^\alpha\prod^N_{\beta=L\atop\beta\ne\alpha}
    \kakko{\Phi^\alpha-\Phi^\beta}}
  e^{iu\kakko{1-\sum^{L-2}_{\gamma=I+1}p^\gamma}\Phi^\alpha}
  \nonumber\\
  &&\times
  \int^{1-\sum^{L-1}_{\gamma=I+1}p^\gamma}_0dp^{L-1}
  e^{iu\kakko{\Phi^{L-1}-\Phi^\alpha}p^{L-1}}
  \nonumber\\
  &&+
  {\kakko{-1}^{N-L+1}\over\prod^N_{\beta=L}\Phi^\beta}
  \int^{1-\sum^{L-2}_{\gamma=I+1}p^\gamma}_0dp^{L-1}
  e^{iu\Phi^{L-1}p^{L-1}}\bigg]
  \nonumber\\
  &=&
  \kakko{iu}^{-N+\kakko{L-1}-1}\Bigg[\sum^N_{\alpha=L}
  {1\over\Phi^\alpha\prod^N_{\beta=L-1\atop\beta\ne\alpha}
    \kakko{\Phi^\alpha-\Phi^\beta}}
  e^{iu\Phi^\alpha\kakko{1-\sum^{L-2}_{\gamma=I+1}}p^\gamma}
  \nonumber\\
  &&+\bigg\{-\sum^N_{\alpha=L}{1\over\Phi^\alpha
    \prod^N_{\beta=L-1\atop\beta\ne\alpha}\kakko{\Phi^\alpha-\Phi^\beta}}
  +{\kakko{-1}^{N-L+1}\over\prod^N_{\beta=L-1}\Phi^\beta}\bigg\}
  e^{iu\Phi^{L-1}\kakko{1-\sum^{L-2}_{\gamma=I+1}p^\gamma}}
  \nonumber\\
  &&+
  {\kakko{-1}^{N-\kakko{L-1}+1}\over\prod^N_{\beta=L-1}\Phi^\beta}\Bigg]
  \nonumber\\
  &=&
  \kakko{iu}^{-N+\kakko{L-1}-1}\Bigg[\sum^N_{\alpha=L}
  {1\over\Phi^\alpha\prod^N_{\beta=L-1\atop\beta\ne\alpha}
    \kakko{\Phi^\alpha-\Phi^\beta}}
  e^{iu\Phi^\alpha\kakko{1-\sum^{L-2}_{\gamma=I+1}}p^\gamma}
  \nonumber\\
  &&+
  {1\over\Phi^{L-1}\prod^N_{\beta=L}\kakko{\Phi^{L-1}-\Phi^\beta}}
  e^{iu\Phi^{L-1}\kakko{1-\sum^{L-2}_{\gamma=I+1}p^\gamma}}
  +{\kakko{-1}^{N-\kakko{L-1}+1}\over\prod^N_{\beta=L-1}\Phi^\beta}\Bigg]
  \nonumber\\
  &=&
  \kakko{-1}^{-N+\kakko{L-1}-1}\Bigg[\sum^N_{\alpha=L-1}
  {1\over\Phi^\alpha\prod^N_{\beta=L-1\atop\beta\ne\alpha}
    \kakko{\Phi^\alpha-\Phi^\beta}}
  e^{-u\Phi^\alpha\kakko{1-\sum^{L-2}_{\gamma=I+1}p^\gamma}}
  \nonumber\\
  &&+
  {\kakko{-1}^{N-\kakko{L-1}+1}\over
    \prod^N_{\beta=L-1}\Phi^\beta}\Bigg]\ .
\end{eqnarray}
(\ref{tsugi}) indicates that (\ref{sekibunkoushiki})
in $L-1$ case holds if it holds in $L$ case.
Thus we conclude that (\ref{sekibunkoushiki}) holds for
$L=N,N-1,\cdots,I+1$.

\section{The Determinant Form of the Exact Result}
\label{sec:det}

Utilizing the relation in appendix \ref{sec:koshiki},
we rewrite the exact result of the trace formula
in the determinant form.

By assigning
\begin{equation}
  k = 1\ ,\
  n = N+1\ ,\
  a_\alpha = e^{-ic_\alpha T}\ ,
\end{equation}
(\ref{app:youso}) becomes
\begin{equation}\label{app:yousogotai}
  \prod^{N+1}_{\beta=1\atop\beta\ne\alpha}
  \kakko{e^{-ic_\alpha T}-e^{-ic_\beta T}}
  =
  \kakko{-1}^{N+1-\alpha}
  {\Delta_1\over\kansu{\tilde\Delta_1}{\alpha}}\ .
\end{equation}
By means of (\ref{app:yousogotai}), we rewrite (\ref{multi:genmitsu}) to
\begin{eqnarray}
  Z &=&
  \sum^{N+1}_{\alpha=1}
  {e^{-iQc_\alpha T}\over\prod^{N+1}_{\beta=1\atop\beta\ne\alpha}
  \left\{1-e^{-i\kakko{c_\beta-c_\alpha}T}\right\}}
  \nonumber\\
  &=&
  \sum^{N+1}_{\alpha=1}
  {e^{-iQc_\alpha T}\over
    e^{iNc_\alpha T}
    \prod^{N+1}_{\beta=1\atop\beta\ne\alpha}
    \kakko{e^{-ic_\alpha T}-e^{-ic_\beta T}}}
  \nonumber\\
  &=&
  \sum^{N+1}_{\alpha=1}
  {e^{-i\kakko{Q+N}c_\alpha T}\over
    \kakko{-1}^{N+1-\alpha}
    {\Delta_1\over\kansu{\tilde\Delta_1}{\alpha}}}
  \nonumber\\
  &=&
  {1\over\Delta_1}
  \sum^{N+1}_{\alpha=1}
  \kakko{-1}^{N+1+\alpha}
  e^{-i\kakko{Q+N}c_\alpha T}
  \nonumber\\
  &&\times
  \left\vert
    \matrix{
      1                         & \cdots &
      1                              &
      1                         & \cdots &
      1                              \cr
      e^{-ic_1T}                & \cdots &
      e^{-ic_{\alpha-1}T}            &
      e^{-ic_{\alpha+1}T}       & \cdots &
      e^{-ic_{N+1}T}                 \cr
      e^{-ic_2T}                & \cdots &
      e^{-i2c_{\alpha-1}T}           &
      e^{-i2c_{\alpha+1}T}      & \cdots &
      e^{-i2c_{N+1}T}                \cr
      \vdots                    & \ddots &
      \vdots                         &
      \vdots                    & \ddots &
      \vdots                         \cr
      e^{-i\kakko{N-1}c_1T}     & \cdots &
      e^{-i\kakko{N-1}c_{\alpha-1}T} &
      e^{-i\kakko{N-1}c_{N+1}T} & \cdots &
      e^{-i\kakko{N-1}c_{N+1}T}      \cr
      }
    \right\vert
    \nonumber\\
    &=&
    {1\over\Delta_1}
    \left\vert
      \matrix{
        1                     & 1                         &
        \cdots                & 1                         \cr
        e^{-ic_1T}            & e^{-ic_2T}                &
        \cdots                & e^{-ic_{N+1}T}            \cr
        \vdots                & \vdots                    &
        \ddots                & \vdots                    \cr
        e^{-i\kakko{N-1}c_1T} & e^{-i\kakko{N-1}c_2T}     &
        \cdots                & e^{-i\kakko{N-1}c_{N+1}T} \cr
        e^{-i\kakko{Q+N}c_1T} & e^{-i\kakko{Q+N}c_2T}     &
        \cdots                & e^{-i\kakko{Q+N}c_{N+1}T} \cr
        }
      \right\vert\ .
\end{eqnarray}
This is the determinant form (\ref{multi:genmitsudet}).

\end{document}